\documentclass[twocolumn,dvipsnames, usenames]{aastex631}
\usepackage[utf8]{inputenc}
\usepackage{amssymb,amsmath,verbatim,mathtools,needspace,enumitem,etoolbox,graphicx,physics,microtype,ulem,breqn}
\usepackage{xspace} 
\usepackage{booktabs}
\usepackage{upgreek} 
\usepackage{acronym}
\usepackage{url}

\newcommand{\MKI}{\affiliation{Department of Physics and Kavli Institute for Astrophysics and Space Research, Massachusetts Institute of Technology, 77 Massachusetts Ave, Cambridge, MA 02139, USA}}
\newcommand{\TTU}{\affiliation{Department of Physics and Astronomy, Box 41051, Texas Tech University, Lubbock, TX 79409}}
\newcommand{\LIGOlabMIT}{\affiliation{LIGO Laboratory, Massachusetts Institute of Technology, 185 Albany St, Cambridge, MA 02139, USA}}

\date{\today}
\submitjournal{ApJ}

\begin{document}

\title{Gamma-Ray Bursts Observed by the Transiting Exoplanet Survey Satellite: 
Prompt Optical Counterparts and Afterglows of Swift-XRT Localized GRBs}

\correspondingauthor{Rahul Jayaraman}
 \email{rjayaram@mit.edu}
\author[0000-0002-7778-3117]{Rahul Jayaraman} \MKI
\author[0000-0002-9113-7162]{Michael Fausnaugh}\MKI\TTU
\author[0000-0003-2058-6662]{George R. Ricker}\MKI
\author[0000-0001-6763-6562]{Roland Vanderspek}\MKI
\author[0000-0001-6331-112X]{Geoffrey Mo}\MKI\LIGOlabMIT

\begin{abstract}

Very few detections have been 
made of optical flashes contemporaneous with prompt high-energy emission 
from a gamma-ray burst (GRB). In this work, we present and analyze light curves 
of GRB-associated optical flashes and afterglows from the Transiting Exoplanet Survey Satellite 
(TESS). Our sample consists of eight GRBs with arcsecond-level localizations from
the X-Ray Telescope on board the Neil Gehrels Swift Observatory.
For each burst, we characterize the prompt optical emission and any observed afterglow, 
and constrain physical parameters for four of these bursts using their TESS 
light curves. This work also presents a straightforward method to correct for TESS’s 
cosmic ray mitigation on 20-second timescales, which allows us to estimate the 
``true'' brightness of optical flashes associated 
with prompt GRB emission. We also highlight TESS's continuous wide-field monitoring capability,
which provides an efficient means of identifying optical emission from GRBs and
characterizing early-time afterglow light curves. Based on empirical detection rates
from Swift and the Fermi Gamma-Ray Space Telescope, up to 10
GRBs per year may fall within the contemporaneous TESS field of view.
\end{abstract}

\section{Introduction}
\label{sec:intro}
Gamma-ray bursts (GRBs) are the most energetic explosions in the universe, releasing
over $10^{51}$\,ergs of energy solely in gamma rays \citep{frail_beaming}. These intense 
transients, typically occurring at cosmological distances, exhibit a bimodal distribution 
in their duration \citep{kouveliotou_grb_populations} that is connected to the nature
of the progenitor. Short GRBs ($\lesssim\,2$\,s) are typically caused by neutron star mergers
(see, e.g., \citealt{eichler_1989,berger_sgrb_review}), 
while long GRBs ($\gtrsim\,10$\,s) canonically result from the core-collapse 
supernovae of low-metallicity stars with stripped 
hydrogen envelopes \citep{woosley_bloom_araa,cano_grb_sn}, but some could
arise from neutron star mergers \citep{rastinejad_kn_lgrb,levan_230307a}. Long
GRBs can often exhibit multiple emission episodes (e.g., \citealt{gendre_ultralong}).

Thousands of GRBs have been detected in the decades since the first observations by
Vela satellites \citep{klebesadel_first_grb}. These detections have been 
enabled by numerous high-energy sky-monitoring
satellites, including the Burst and Transient Source Experiment 
(BATSE) onboard the Compton Gamma-Ray Observatory (e.g., \citealt{batse_catalog}),
the BeppoSAX mission \citep{bepposax_grb},
the High-Energy Transient Explorer-2 \citep{hete_paper}, the Neil Gehrels 
Swift Observatory \citep{swift_mission}, and the Fermi Gamma-Ray Space Telescope 
\citep{fermi_gbm,fermi_lat}. 

GRBs have been confidently associated with rapidly-decaying optical transients 
(``afterglows'') following the discovery of one associated with GRB 970228 
\citep{van_paradijs_afterglow,wijers_afterglow_1997}. 
Afterglows arise from the 
cooling and expanding ejecta shocking the circum-burst medium
\citep{meszaros_afterglow,sari_piran_narayan}. Since GRB 970228,
$\sim$\,1000 optical afterglows have been detected.\footnote{An
unofficial catalog of well-localized GRBs can be found at \\ \url{https://www.mpe.mpg.de/~jcg/grbgen.html},
along with information about afterglow detections in X-ray, optical, and radio.} 
Since the 1990s, the identification of prompt optical counterparts and afterglows
has been enabled by the rapid distribution of GRB localizations via the 
General Coordinates Network (GCN).\footnote{\url{https://gcn.nasa.gov/circulars}}.

While the detection of numerous optical afterglows has shed light on the diversity 
of GRBs and the evolution of their ejecta at late times, it has been more difficult 
to identify optical counterparts contemporaneous with the high-energy emission from the GRB. 
Such episodes are often referred to as ``prompt'' optical emission. 
Searches for prompt emission began in the 1990s, with early upper 
limits on the prompt optical flux established by the Gamma-Ray Optical Counterpart 
Search Experiment and its successor, the Livermore Optical Transient Imaging 
System (LOTIS; \citealt{park_grocse,lotis_paper}). The first detection of prompt 
emission, in GRB 990123, was made using the Robotic Optical Transient Search 
Experiment \citep{akerlof_1999_prompt}. The next detection occurred five 
years later: Prompt optical emission from GRB 041219A 
\citep{vestrand_2005_prompt,blake_ir_041219a} was 
detected by both the RAPTOR\footnote{RAPid Telescopes for 
Optical Response} network \citep{raptor} and the 
Peters Automated Infrared Imaging Telescope \citep{bloom_paritel}. RAPTOR also detected prompt
emission from GRB 050820a \citep{vestrand_2006_energy}.

In the 2000s, several more optical counterparts were found by 
LOTIS \citep{superlotis_results}. Concurrently, the TAROT\footnote{T\'elescopes \`a 
Action Rapide pour les Objets Transitoires} network \citep{tarot_paper} placed 
upper limits on prompt optical flux from other GRBs and detected prompt emission
from GRB 060111B \citep{060111B_prompt}. Early-time optical observations of 
GRB 080319B \citep{racusin_nakedeye_correlation} by
Pi of the Sky \citep{pi_of_sky} and 
TORTORA\footnote{Telescopio Ottimizzato per la Ricerca dei Transienti Ottici 
RApidi} \citep{tortora} showed a clear distinction between prompt emission, 
reverse shock, and afterglow in the light curve. Since GRB 080319B, 
only a few prompt optical counterparts have been detected \citep{vestrand_130427a,troja_prompt_polarization,zhang_2018_prompt,becerra_prompt_2021,xin_2023_prompt,oganesyan_prompt_2023}. 

The physical mechanism underlying prompt optical emission
remains poorly understood. \citet{mezaros_rees_internal_shock}
suggested that the prompt optical flash from GRB 990123 arose from
internal shocks, or a reverse shock from blast wave generated by the 
GRB interacting with the circum-burst medium (also 
discussed in \citealt{2000ApJ...545..807K}). Modeling of the reverse shock hypothesis
\citep{meszaros_afterglow,sari_piran_prompt} found that a reverse shock 
would have an energy comparable to the forward shock, but 
radiate at optical wavelengths. Other possibilities for the prompt optical emission
include synchrotron self-Compton
upscattering \citep{panaitescu_kumar_unified_prompt} and inverse Compton scattering in
the thermal plasma behind the forward shock \citep{inverse_compton_optical_flash}.

In an ensemble analysis of prompt optical emission from
GRBs, \citet{kopac_prompt_sample_2013} suggested that synchrotron radiation from
shocks internal to the jet could explain the observed optical flashes.
\citet{oganesyan_sample_synchrotron} analyzed a sample of 
21 GRBs, some with both prompt optical observations and early-time data from the Swift X-Ray
Telescope (Swift-XRT); half of their sample had
gamma-ray through optical spectra consistent with synchrotron emission.

The paucity of prompt optical detections of GRBs reflects the difficulty of
detecting such emission serendipitously from the ground. Telescopes must
be observing the particular region of sky in which a GRB occurs at 
the time of trigger (e.g., \citealt{xin_2023_prompt}),
or rapidly slew to tile the initial localization of the GRB (e.g.,
\citealt{akerlof_1999_prompt,vestrand_2005_prompt}). These issues can be 
mitigated by space-based optical observatories
with large fields of view, which can continuously monitor emissions
from the minutes to hours before the explosion,
through the time of the gamma-ray burst trigger, and after the cessation of
prompt gamma-ray emission. Such continuous observations allow for 
the identification of prompt optical counterparts, the afterglow, and late-time emission such as 
kilonovae \citep{gw170817_paper} or supernovae \citep{galama_sn}. 
These observations could also constrain the existence of 
optical precursors, which have never been observed
\citep{blake_bloom_precursors,precursor_080319B}, but may relate to the gamma-ray
precursors that are seen in $\sim$\,20\% of bursts \citep{koshut_precursor_1995, 
lazzati_batse,swift_precursor,fermi_precursor}.

In this work, we searched for optical emission from 22 GRBs with arcsecond-level 
localizations from Swift-XRT in data from the Transiting
Exoplanet Survey Satellite (TESS). Of these, we detected 
nine bursts with optical emission.
Section \ref{sec:obs} discusses the TESS observations, high energy data,
and our searches for optical emission.
Section \ref{sec:grb_lcs} 
presents our power-law fits to the afterglows and constraints on 
the brightness of optical flashes at the time of the GRB trigger. 
Section \ref{sec:physical_params} interprets the light curves and
constrains burst parameters.
Finally, Section \ref{sec:disc} discusses the utility of TESS for GRB science.


\begin{rotatetable*}
\vspace{-0.7in}
\begin{deluxetable*}{lrrrrrrrrrrrrr}
\tablecaption{\small GRBs with Swift-XRT localizations falling within the TESS FOV at the 
time of trigger. These GRBs occurred between the start of the TESS mission (in 2018 July)
and 2023 September, corresponding to all of TESS Sectors 1--69.
The nine GRBs with detections in TESS are highlighted in bold. 
BTJD (Barycentric TESS Julian Day) is defined as JD-TDB--2\,457\,000, and is measured
at the solar system barycenter. JD-TDB is the Barycentric Dynamical Time standard
as detailed in \citet{2010PASP..122..935E}. GRB
coordinates are obtained from the Swift-XRT GRB Catalogue (\url{https://www.swift.ac.uk/xrt_live_cat/}), 
which contains analyses based on \citet{swift_xrt_grbs}. 
For all GRBs, we note the 3-$\sigma$ detection threshold, calculated
based on the rms scatter in the light curve before the trigger.
Entries with -- indicate that there was no
information available for the T$_{\rm 90}$ parameter, or that
the GRB occurred during an observing gap in TESS, precluding a
calculation of the limiting magnitude.
$E_{B-V}$ extinction values, calculated using the reddening maps from
\citet{schlafly_finkbeiner_extinction} and the coefficients from \citet{cardelli_extinction},
account only for Galactic extinction. The measured flux
was divided by the extinction correction factor (enumerated in 
the last column) to estimate the unattenuated
flux values; details of this
calculation are provided in Section \ref{subsec:tess-info}. Horizontal lines
indicate changes in the TESS FFI cadence, from 1800\,s to 600\,s, and from
600\,s to 200\,s.}
\label{tab:tess_grb_info}
\tablehead{\multicolumn{1}{c}{Identifier}& 
\multicolumn{6}{c}{\textbf{Coordinates (J2000)}} &
\multicolumn{2}{c}{\textbf{Trigger}} &
\multicolumn{2}{c}{\textbf{GRB Properties}} &
\multicolumn{1}{c}{3-$\sigma$ Limit} &
\multicolumn{1}{c}{$E_{B-V}$} & 
\multicolumn{1}{c}{Extinction}\\
\multicolumn{1}{c}{}&
\multicolumn{3}{c}{Right Ascension} &
\multicolumn{3}{c}{Declination} &
\multicolumn{1}{c}{BTJD} &
\multicolumn{1}{c}{Sector} &
\multicolumn{1}{c}{T$_{\rm 90}$ (s)} &
\multicolumn{1}{c}{$\sigma$-T$_{\rm 90}$} &
\multicolumn{1}{c}{(T$_{\rm mag}$)} & 
\multicolumn{1}{c}{(mag)} & 
\multicolumn{1}{c}{Correction}}
\startdata
GRB 180727A & 23h & 06m & 39.84s & --63d & 03m & 06.48s & 1327.09850 & 1 & 1.1$^a$ & 0.2 & 18.74 & 0.019 &  0.965 \\
GRB 180924A & 03h & 16m & 47.87s & --58d & 31m & 57.00s & 1386.14331 & 3 & 95.1$^b$ & 10.9 & 19.01 & 0.025 &  0.955 \\
GRB 181022A & 03h & 47m & 09.94s & --29d & 22m & 56.28s & 1414.23386 & 4 & 6.74$^c$ & 2.30 & 18.98 & 0.012 &  0.979 \\
GRB 190422A & 12h & 08m & 08.42s & --60d & 13m & 27.48s & 1596.45954 & 10 & 213.25$^d$ & 10.75 & 17.98 & 1.402 &  0.089 \\
GRB 190630C & 19h & 35m & 31.25s & --32d & 44m & 38.04s & 1665.50153 & 13 & 38.4$^e$ & 9.3 & 17.88 & 0.099 &  0.833 \\
\textbf{GRB 191016A}$^f$ & 02h & 01m & 04.65s & +24d & 30m & 35.68s & 1772.67919 & 17 & 219.70$^g$ & 183.35 & 18.76 & 0.094 &  0.841 \\
GRB 200303A & 14h & 10m & 52.54s & +51d & 21m & 33.37s & 1911.61090 & 22 & 94.2$^h$ & 6.4 & 19.08 & 0.014 &  0.975 \\
GRB 200324A & 14h & 50m & 41.71s & +35d & 56m & 29.98s & 1933.19817 & 23 & --$^i$ & -- & 18.87 & 0.012 &  0.977 \\
\textbf{GRB 200412B} & 18h & 33m & 15.19s & +62d & 31m & 57.00s & 1951.88161 & 23 & 6.08$^d$ & 0.29 & 18.88 & 0.050 &  0.911 \\
\hline
\textbf{GRB 200901A} & 04h & 07m & 07.96s & --59d & 53m & 26.88s & 2093.65995 & 29 & 20.37$^j$ & 7.55 & 18.67 & 0.014 &  0.975 \\
\textbf{GRB 210204A}$^k$ & 07h & 48m & 19.34s & +11d & 24m & 33.98s & 2249.77660 & 34 & 206.85$^d$ & 2.29 & 18.19 & 0.028 &  0.949 \\
GRB 210222B & 10h & 18m & 25.37s & --14d & 55m & 54.12s & 2268.44869 & 35 & 12.82$^l$ & 1.28 & -- & 0.089 &  0.848 \\
GRB 210419A & 05h & 47m & 24.18s & --65d & 30m & 07.92s & 2323.78791 & 37 & 64.43$^m$ & 11.69 & $\sim$17.5 & 0.066 &  0.884 \\
\textbf{GRB 210504A} & 14h & 49m & 33.84s & --30d & 32m & 00.24s & 2339.08622 & 38 & 135.06$^n$ & 9.57 & 18.24 & 0.122 &  0.798 \\
GRB 210730A & 09h & 58m & 22.08s & +69d & 41m & 22.09s & 2425.70386 & 41 & 3.86$^o$ & 0.66 & 17.66 & 0.074 &  0.872 \\
GRB 220319A & 14h & 32m & 53.86s & +61d & 17m & 43.30s & 2658.23949 & 49 & 6.44$^p$ & 1.54 & 18.19 & 0.009 &  0.983 \\
\textbf{GRB 220623A} & 09h & 41m & 34.49s & +75d & 49m & 15.60s & 2753.79238 & 53 & 57.11$^q$ & 8.53 & 18.16 & 0.032 &  0.942 \\
GRB 220708A & 06h & 52m & 30.17s & +72d & 08m & 28.00s & 2768.69110 & 53 & 4.4$^r$ & 1.0 & 17.99 & 0.152 &  0.755 \\
\hline
GRB 221120A & 02h & 45m & 21.56s & +43d & 14m & 35.27s & 2904.40134 & 58 & 0.79$^s$ & 0.16 & 17.61 & 0.082 &  0.860 \\
\textbf{GRB 230116D} & 06h & 34m & 28.28s & +49d & 52m & 22.30s & 2961.38386 & 60 & 41.00$^t$ & 11.18 & 17.84 & 0.100 &  0.832 \\
\textbf{GRB 230307A}$^u$ & 04h & 03m & 26.24s & --75d & 22m & 43.68s & 3011.15549 & 62 & 34.56$^d$ & 0.57 & 17.87 & 0.077 &  0.867 \\
\textbf{GRB 230903A} & 00h & 39m & 38.81s & --40d & 56m & 57.48s & 3191.22939 & 69 & 2.54$^v$ & 0.27 & 17.83 & 0.011 &  0.980 \\
\enddata
\tablecomments{(a) \citet{180727A_swift_bat_analysis_gcn}. (b) \citet{180924A_swift_bat_analysis_gcn}. (c) \citet{181022A_swift_bat_analysis_gcn}.
(d) Fermi GBM Burst Catalog (FERMIGBRST; \citealt{fermi_grb_cat_1, fermi_grb_cat_2, fermi_grb_cat_4, fermi_grb_cat_3}. 
(e) \citet{190630C_swift_bat_analysis_gcn}. (f) Analyzed and discussed in \citet{smith_191016a}. 
(g) \citet{191016A_swift_bat_analysis_gcn}. (h) \citet{200303A_swift_bat_analysis_gcn}. (i) \citet{200324A_swift_bat_analysis_gcn}. 
(j) \citet{200901A_swift_bat_analysis_gcn}. (k) This GRB was associated with the ZTF-triggered optical transient 
AT2021buv/ ZTFaagwbjr, discovered with ZTFReST \citep{andreoni_ztfrest}. The ZTF light curve was analyzed in \citet{kumar_afterglow_flares} and 
\citet{ho_dirty_fireballs}. (l) \citet{2102222B_swift_bat_analysis_gcn}. (m) \citet{210419A_swift_bat_analysis_gcn}.
(n) \citet{swift_bat_210504a}. (o) \citet{210730A_bat_analysis_gcn}. (p) \citet{220319A_swift_bat_analysis_gcn}. (q) \citet{220623a_refined}. 
(r) \citet{220708A_swift_bat_analysis}. (s) \citet{221120A_bat_refined_analysis}. (t) \citet{swift_bat_230116d}. 
(u) \citet{fausnaugh_230307a,levan_230307a,yang_230307a}. (v) \citet{230903a_swift_bat_analysis}.}
\end{deluxetable*}
\end{rotatetable*} 

\section{TESS observations of GRBs}
\label{sec:obs}

TESS is an all-sky survey whose primary aim is to detect
transiting exoplanets orbiting bright stars \citep{ricker_tess}. 
Launched in 2018 April, TESS observes 
a given 2304 deg$^2$ region of the sky for $\sim$28 days at a time (a period known as a
``sector'').\footnote{Information about 
TESS pointings is available at \url{https://tess.mit.edu/observations/}.} 
During the first five years of the mission, TESS observed over 90\% of the sky,
and revisited fields roughly every other year. 
TESS's CCD detectors have a plate scale of 21''/pixel.

During its Prime Mission, from 2018 July to 2020 July (Sectors 1--26), 
TESS obtained images of its entire field of view (``full-frame images''; 
FFIs) at a cadence of 1800\,s.
During TESS's Extended Mission 1, from 2020 July to 2022 September (Sectors 27--55), 
the FFI exposure times were shortened to 600\,s. Starting in 2022 September,
TESS's Extended Mission 2 (Sectors 56--83) further 
reduced the FFI cadence to 200\,s and added weekly data downlinks---making TESS 
much more useful for timely follow-up of fast transients such as GRBs.

TESS's ability to continuously monitor over 2\,000\,deg$^2$ of the sky for at least a
month at a time is useful for detecting prompt optical flashes from GRBs and 
other associated emission. 
TESS data have already been used to study individual GRBs and their 
afterglows: \citet{smith_191016a} analyzed the afterglow of GRB 191016A; \citet{fausnaugh_230307a} 
presented the TESS light curve of GRB 230307A, with a prompt flash
preceding a fainter afterglow; and \citet{roxburgh_afterglows}
identified four GRB afterglows in TESS data.

\subsection{Sample Selection}
We analyzed GRBs from the Swift-XRT catalog 
\citep{swift_xrt_grbs}\footnote{\url{https://www.swift.ac.uk/xrt_live_cat/}}---which 
includes GRBs localized to within $\lesssim$10'' \citep{swift_xrt_positions}---and
identified 22 GRBs where the trigger time
and localization coincided with TESS observations of that sky region. Our search spanned the first five years
of the TESS mission---from the start of Sector 1 in 2018 July, to the end of TESS Sector 69, 
in 2023 September. An additional 39 bursts were detected by Swift-BAT over the time 
period in consideration that lacked a corresponding XRT detection. None
of these arcminute-level localizations
fell within the contemporaneous TESS field of view.
The list of TESS-coincident GRBs, including coordinates,
trigger times, and GRB properties, is given in Table \ref{tab:tess_grb_info}.

\subsection {TESS Light Curves}
\label{subsec:tess-info}

\begin{table*}
\caption{TESS light curves of the eight GRBs analyzed in this work. The
light curves start 0.5\,d before the trigger, and end 1.5\,d post-trigger. We
report the flux in counts, magnitudes, and provide the $F_\nu$ (in Jy), along with these values'
associated uncertainties. We also include the background estimate from our
photometry routine. For cadences with non-detections,
we include the limiting $T_{\rm mag}$ and a -- in the Flux and Uncertainty columns.}
\centering
\begin{tabular}{rcccccccc}
\hline
\hline
\multicolumn{1}{c}{GRB Identifier} &
\multicolumn{1}{c}{BTJD} &
\multicolumn{1}{c}{Diff. Flux} &
\multicolumn{1}{c}{Uncertainty} &
\multicolumn{1}{c}{T$_{\rm mag}$} &
\multicolumn{1}{c}{Uncertainty} &
\multicolumn{1}{c}{Flux} &
\multicolumn{1}{c}{Uncertainty} &
\multicolumn{1}{c}{Background} \\
\multicolumn{1}{c}{} &
\multicolumn{1}{c}{(BJD--2\,457\,000)} &
\multicolumn{2}{c}{(ct\,s$^{-1}$)} &
\multicolumn{1}{c}{} &
\multicolumn{1}{c}{} &
\multicolumn{2}{c}{(mJy)} &
\multicolumn{1}{c}{(ct s$^{-1}$)} \\
\hline 
200412B & 1951.38554 & -0.777 & 0.564 & 18.880 & -- & -- & -- & 1.298 \\
200412B & 1951.40637 & 0.939 & 0.559 & 18.880 & -- & -- & -- & 0.656 \\
200412B & 1951.42721 & -1.098 & 0.555 & 18.880 & -- & -- & -- & 0.146 \\
200412B & 1951.44804 & -1.418 & 0.554 & 18.880 & -- & -- & -- & 0.065 \\
200412B & 1951.46888 & -1.194 & 0.564 & 18.880 & -- & -- & -- & 0.713 \\
200412B & 1951.48971 & -0.715 & 0.551 & 18.880 & -- & -- & -- & 0.445 \\
200412B & 1951.51054 & 1.663 & 0.549 & 18.787 & 0.340 & 0.079 & 0.029 & 2.166 \\
... & ... & ... & ... & ... & ... & ... & ... & ... \\
\hline
\end{tabular}
\label{tab:lc}
\tablecomments{The entirety of this table is available in machine-readable
format.}
\end{table*}

We extract TESS light curves of GRBs using the FFIs generated by the
TESS Image CAlibrator (TICA; \citealt{fausnaugh_tica}). We first build a reference
image from the median of 20 FFIs that have a low background, and then subtract 
the reference from the other FFIs using the ISIS image subtraction package 
\citep{isis}. Then, for each GRB, we perform forced photometry on
the generated difference images at the GRB coordinates. Both the difference 
imaging and photometry procedures are detailed in
\citet{fausnaugh_diff_imaging,fausnaugh_ia_2023}.
The observed counts are then converted to TESS Vega-magnitudes ($T$), 
and then to Jy ($F_\nu$).
To do so, we calculated the Vega zeropoint in the TESS bandpass using
the CALSPEC Vega model \citep{2014AJ....147..127B}, as implemented in 
the synthetic photometry package {\tt synphot} 
\citep{synphot}. We found that a TESS Vega-magnitude of $T=0$ 
corresponds to a flux of 2583\,Jy. Thus, the equations to calculate
$T$\footnote{The 
TESS ``zeropoint'' of 20.44, corresponding to a detector count rate of 1\,e$^-$\,s$^{-1}$, 
has an uncertainty of 0.05 that propagates
to all our magnitude estimates \citep{fausnaugh_diff_imaging}. This was calculated
from the equivalence in the TESS instrument handbook (linked in footnote 10) that 
15\,000\,e$^-$\,s$^{-1}$ corresponds to $T=10$.} and $F_\nu$ are
$T = -2.5\,\log_{10}\left(\frac{N}{t \times 0.99 \times 0.8}\right) + 20.44$,
and $F_\nu = 2583\,{\rm Jy} \times 10^{-T/2.5}$, respectively.
respectively. Here, $N$ is the total observed counts from a source in a given FFI, 
and $t$ is the observational cadence (exposure time).
The factors of 0.99 and 0.8 arise from the frame transfer efficiency and the 
cosmic-ray mitigation process,
respectively. Cosmic-ray mitigation is further discussed in Section
\ref{subsec:prompt_analysis_details} and Appendix \ref{app:crm}. 
In Table \ref{tab:lc}, we present light curves spanning from 0.5\,d
pre-trigger to 1.5\,d post-trigger for the eight GRBs from Swift-XRT with confident
detections in TESS; the light curves in the Table include both differential counts
and $F_\nu$ (given in Jy). These are flux-calibrated, i.e., the light curve has been
shifted to ensure a baseline of 0 counts by subtracting
the mean value of the out-of-burst light curve (assuming a
constant background term).

We then corrected for Galactic extinction as follows: First, we assumed an intrinsic power-law
spectrum for the GRB of the form $F_0 (\nu) \propto \nu^{-1}$, 
and calculated the response in the TESS bandpass\footnote{\url{https://heasarc.gsfc.nasa.gov/docs/tess/data/tess-response-function-v2.0.csv}}
using {\tt synphot}. We then calculated the extinction of the input spectrum in the TESS bandpass using a 
\citet{cardelli_extinction} extinction law (with $R_V = 3.1$, which is handled by  the {\tt mwavg} 
option in {\tt synphot}), and $E_{B-V}$ values from 
\citet{schlafly_finkbeiner_extinction}.\footnote{These are from the online extinction
calculator hosted by the NASA/IPAC Extragalactic Database (\url{https://ned.ipac.caltech.edu/extinction_calculator})}.
The relationship between the extinction in 
the TESS bandpass and the $V$ band (from \citealt{landolt_v}) is given by:
$A_{\rm TESS} = (-0.0097\pm0.0475)\,{A_V}^2 +\,(0.6470\pm0.3047)\,A_V +\,(0.0025\pm0.4087)$
We also calculated the conversion from TESS Vega magnitudes to AB magnitudes:
$T_{\rm AB} = (0.3697\pm0.0001) + T_{\rm Vega}$.
Finally, we recalculated the 
response in the TESS bandpass of the extinguished spectrum; the ratio of the unextinguished to extinguished 
flux yields the correction factor for the observed GRB light curves,
given in the rightmost column of Table \ref{tab:tess_grb_info}. The column to 
its left gives the $E_{B-V}$ values used in this calculation.


After correcting for extinction, we then calculated the monochromatic flux 
$\nu F_{\nu}$, where $\nu$ 
is the frequency corresponding to the pivot wavelength of the TESS bandpass (784 nm),
calculated using {\tt synphot}.
All flux values reported throughout this work are in $\nu F_{\nu}$ (with physical units of 
erg\,cm$^{-2}$\,s$^{-1}$), where $\nu$ is the corresponding central frequency of the TESS
bandpass, $3.824\times10^{14}$\,Hz. We
define the Barycentric TESS Julian Day (BTJD) as JD-TDB--2\,457\,000, where
JD-TDB is a timesystem in which times are measured at the solar system 
barycenter \citep{2010PASP..122..935E}. Further information on timing corrections 
is given in Section \ref{subsubsec:time_systems}.


\subsection{Identifying GRB Signals in TESS}
\label{subsec:identifying_signals}

\begin{figure*}
\centering
    \includegraphics[width=\textwidth]{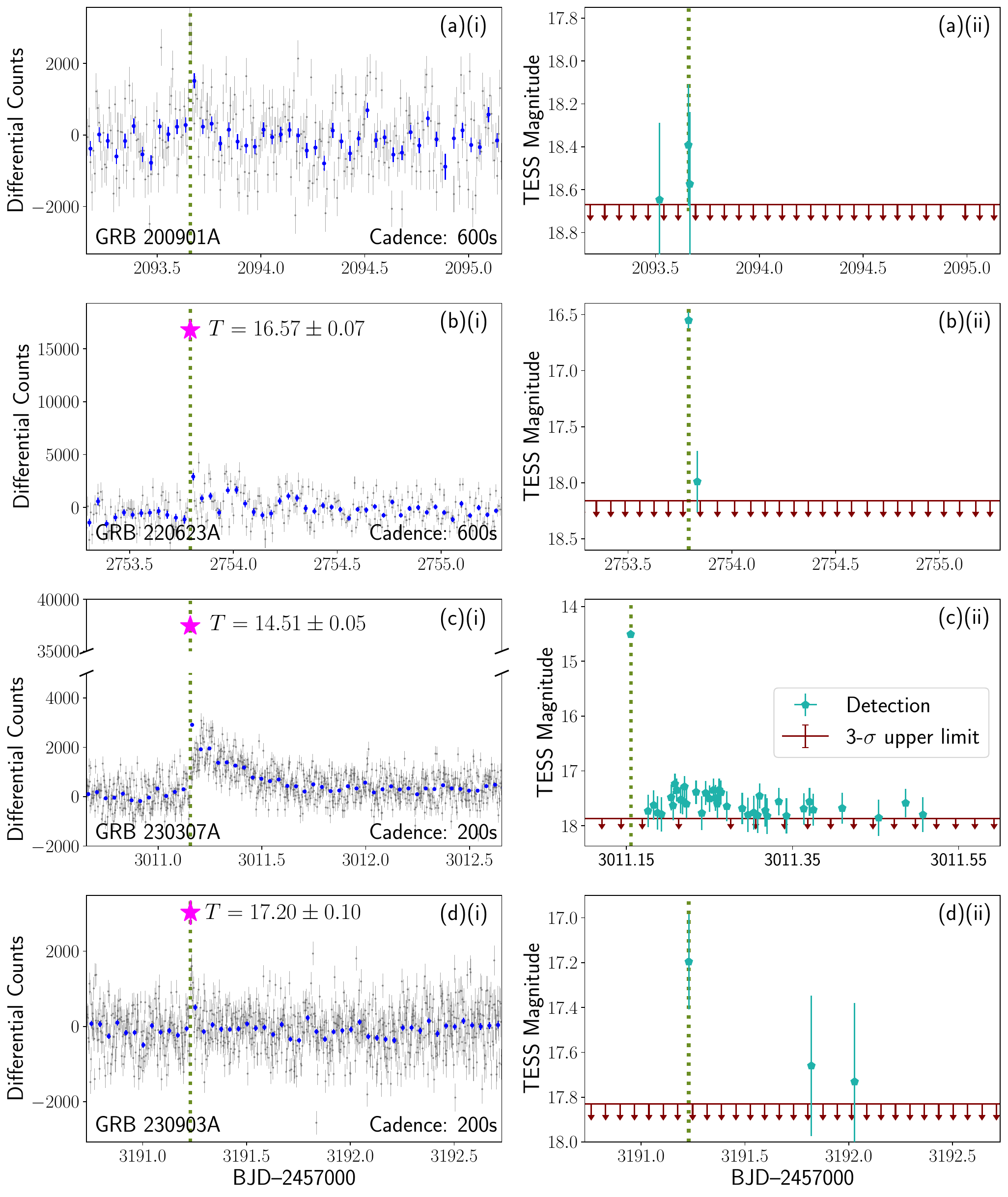}
    \caption{TESS light curves of Swift-XRT localized GRBs with prompt emission detected at the 3-$\sigma$
    level in the TESS difference images (see section \ref{subsec:identifying_signals} for
    further information). The left panel shows the flux in raw counts, as measured in 
    the difference images; zero is the flux in the reference image 
    (constructed using the procedure described in \citealt{fausnaugh_diff_imaging,fausnaugh_ia_2023}). 
    The flux at the native cadence is shown in light gray, while the light curve binned to 1\,hr
    is shown in blue. The right
    panel shows the light curve in magnitudes---the pentagons correspond to 3-$\sigma$
    detections of a source, while the arrows represent upper limits on 
    the flux. Magnitudes were calculated by shifting the observed light curve (in counts) so that the out-of-burst
    portion was at zero flux; uncertainties were rescaled to be consistent with the rms scatter. The time of trigger is indicated by the green dotted line, and likely
    detections of prompt emission are indicated by a pink star in the left panels.}
    \label{fig:prompt_count_mags}
\end{figure*}

\begin{figure*}
\centering
    \includegraphics[width=\textwidth]{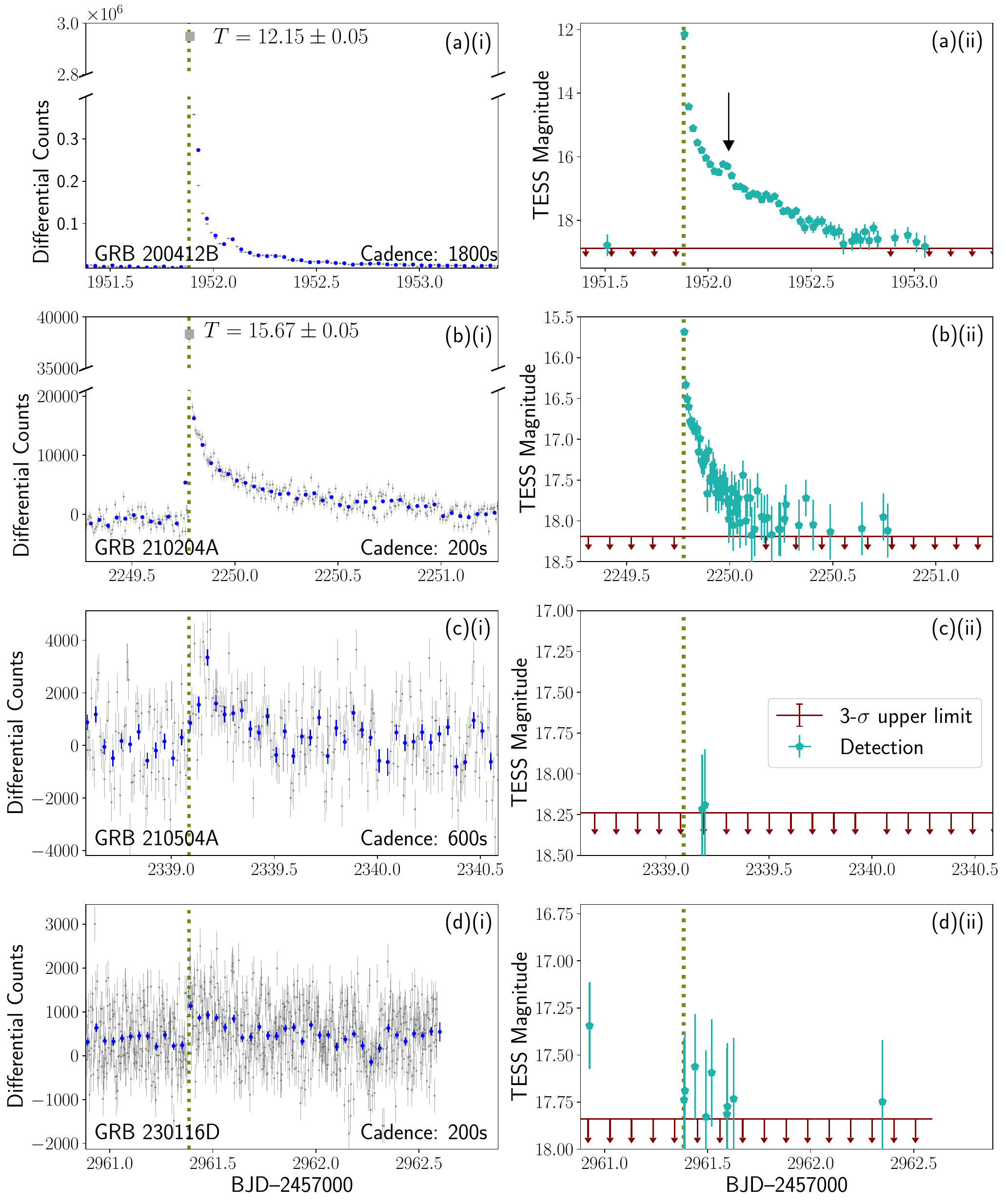}
    \caption{As Figure \ref{fig:prompt_count_mags}, but for bursts without a 
    clear detection of prompt emission in a single cadence. This figure also includes lower-significance
    detections of possible afterglows, such as GRBs 210504A and 230116D in panels (c) and (d), respectively. The gray
    squares in panels (a)(i) and (b)(i) represent detections that likely have some contribution from
    prompt emission, due to the longer cadence FFIs (1800\,s and 600\,s, respectively).
    The arrow in the top right panel
    indicates the flare observed in the afterglow of GRB 200412B.}
    \label{fig:afterglow_count_mags}
\end{figure*}

After flux-calibrating the TESS light curves for our sample of 22 GRBs, we looked for signatures of prompt and 
afterglow emission by both visual inspection and searching for any cadences exhibiting a $>$3-$\sigma$ flux excess 
within $\pm$0.2\,d of the GRB trigger; this excess was calculated relative to the rms scatter of the light curve 
baseline before the trigger, using a window of 0.5\,d in a flat portion of the light curve. 
We identified such an excess for 9 GRBs; we chose to exclude GRB 191016A
from our sample, as it had been analyzed in \citet{smith_191016a}. Figures \ref{fig:prompt_count_mags} and 
\ref{fig:afterglow_count_mags}
show light curves in both counts and magnitude for the other 8 GRBs.


Some notes about the data for specific GRBs follow. 

\begin{itemize}

\item GRB 200901A could be contaminated by the flux from a 
$T=15.6$ source within  30'' (1.5 TESS pixels) 
of the GRB position (TIC 358253511)

\item GRB 210222B occurred in the observing gap during data downlink 
in Sector 35; hence, there is 
no data at the time of burst. However, we still searched for afterglow emission from 
this GRB in TESS, given the confident afterglow detections reported via GCN 
(e.g., \citealt{210222b_lco_afterglow,2021GCN.29546....1H}). 
This GRB also occurred less than 3 TESS pixels (1 arcmin) away from a known variable star 
TIC 385212568, with a magnitude of $T=11.4$.

\item GRB 210419A suffers significant contamination from a nearby variable star (TIC 149597885; 7'' sky separation, or $\lesssim$0.5 TESS pixels) with magnitude $T=15.9$. 
The variability in this star 
is clearly visible in light curves from previous sectors.

\item Data for GRB 220623A were affected by time-varying backgrounds from Earthshine 
around and after the time of trigger.

\item GRB 230116D is less than 3 TESS pixels (60'') away from a $T = 13.5$ magnitude star, TIC\,253030410.
Other nearby faint stars (which have $T \sim 18$) may 
also add noise to the light curve.

\end{itemize}

We also note that the 1800\,s cadence during TESS's prime mission and,
to some extent, the 600\,s cadence during the Extended Mission 1, may
obscure the different phases of GRB emission. For instance, an 1800\,s 
FFI spanning the time of trigger may contain contributions from the 
prompt emission and the early afterglow. For GRBs where this is the case,
we attempted to remove the afterglow's contribution to the fluence; more
details are given in the text describing the analysis for each GRB.

\subsection{Analysis of Prompt Emission}
\label{subsec:prompt_analysis_details}
The TESS FFI exposure times throughout the mission of 1800\,s, 600\,s, 
and 200\,s, are are typically
much longer than the prompt gamma-ray emission from a burst (see, e.g.,
Fig. 1 from \citealt{tarnopolski_t90_dist}). Consequently,
if the prompt optical emission were to have the same temporal profile as the
high-energy emission (as was observed in \citealt{vestrand_2005_prompt,vestrand_130427a}
and \citealt{racusin_nakedeye_correlation}, for photons between 15--350 keV), 
the flux measured by TESS would underestimate the actual flux, 
due to the exposure time being considerably longer than the
duration of the burst itself. The magnitude of this 
effect can be estimated from the ratio of the GRB duration to the FFI exposure time.
The GRB's duration is estimated typically with the $T_{\rm 90}$ parameter, the 
time interval during which 90\% of its prompt high-energy flux is emitted.

An additional effect which can reduce the number of detected counts
seen by TESS for GRBs is the on-board cosmic ray rejection algorithm.
This algorithm, described in detail in Appendix \ref{app:crm},
attempts to mitigate the impact of cosmic rays by removing outlier
pixel values at the native 2\,s CCD integration time, before they
are co-added to create the FFIs. As a result, this procedure
may affect observations of rapid variability on timescales 
of a few seconds. A full description of the TESS cosmic-ray mitigation (CRM) 
technique is given in $\S$5.1 of the TESS Instrument Handbook.\footnote{
\url{https://archive.stsci.edu/files/live/sites/mast/files/home/missions-and-data/active-missions/tess/_documents/TESS_Instrument_Handbook_v0.1.pdf})} Typically,
the fluence is underestimated by 20--25\% due to the effects of CRM; this value
can be much larger for particularly bright bursts. However, for most bursts, CRM has no net effect on
the calculated magnitude across the entirety of the exposure, as the increase 
in counts is roughly offset by the 25\% increase in the exposure time.

In order to characterize the effect of TESS CRM on our sample of bursts, we compared the optical
data to the high-energy light curve from Fermi-GBM and/or Swift-BAT under 
the assumption that the optical light curve has the same shape as the high-energy
light curve.

\subsubsection{High-energy observations}

We downloaded the data products available for the eight bursts of interest from the Fermi-GBM 
and Swift-BAT catalogs at NASA's High Energy Astrophysics Science Archive Research Center 
(HEASARC)\footnote{\url{https://heasarc.gsfc.nasa.gov/}}. For the Fermi data, we used the
Time-Tagged Event (TTE) data from the single NaI detector with the highest signal-to-noise detection
of the GRB. For Swift-BAT data (used to analyze GRBs 220623A and 230903A),
we used the ``Rate'' light curves, which are sampled at 64\,ms. All gamma-ray light curves were 
corrected for a constant background that was calculated from the data in the 100\,s after gamma-ray
emission activity ceased.

All binning and analyses were done using either the Fermi-GBM Data Tools
\citep{GbmDataTools}, based on the Multi-Mission Maximum
Likelihood (3ML) framework \citep{3ML_code}, or the 
{\tt swift\_too}\footnote{\url{https://www.swift.psu.edu/too_api/index.php}}
package.

\subsubsection{Ensuring consistency in time systems}
\label{subsubsec:time_systems}
Light travel time effects between Fermi, Swift, and TESS make it
imperative to ensure that time systems are consistent when comparing data between
these missions. We require 1\,s accuracy on our light curves to compare Fermi and
Swift light curves to individual frame readouts by TESS, which occur every 2\,s. 
Given that the light travel time from TESS to the Earth can be up to 1.2\,s at
apogee, we corrected all time stamps to the JD-TDB time system \citep{2010PASP..122..935E},
using {\tt astropy} to account for the light travel time to the solar system barycenter.
For this analysis, we assumed that Fermi and Swift are located at the geocenter;
these satellites are actually in low Earth orbit, however, so this assumption introduces 
a $\sim$23\,ms error into our analysis. However, this discrepancy
is small compared to the 1\,s accuracy that is required to compare to the TESS FFIs.

To ensure an accurate barycentric correction for the TESS data at the time of 
trigger, we determined the exact position of TESS during the GRB using 
{\tt SpiceyPy} \citep{spiceypy}, a Python wrapper to NASA's SPICE system
\citep{NAIF-spice,NAIF-spice-geometry}. The correction was then applied using
the coordinates of the GRB, as derived from Swift-XRT. While there is a barycentric
correction value provided as part of the full-frame images produced by the 
Science Processing Operations Center
pipeline \citep{jenkins-spoc}, it was unsuitable for
our purposes, as this correction is calculated at the center of the field, which
introduces a systematic error of a few seconds for sources that fell elsewhere
on the detector.

We note that there may be some offset between the trigger times reported
in BTJD and the times reported in UT (Universal Time), when converted to JD, throughout Section
\ref{sec:grb_lcs}. This discrepancy primarily arises from the differing path
lengths of light from the GRB to the Earth, where UT is measured, and to the 
solar system barycenter, where TDB is measured. Values for these corrections
for the eight GRBs analyzed in this paper
will be provided upon reasonable request to the corresponding author.


\section{Optical Light Curve Analysis}
\label{sec:grb_lcs}

\begin{table*}
\caption{
Best-fit parameters for the six Swift-XRT GRBs detected by TESS exhibiting a clear afterglow.
We fit power laws (broken and single, as in Equations \ref{eqn:bpl} and \ref{eqn:spl}, 
respectively) to the extinction-corrected light curves.
The power-law type is indicated in the second 
column. We give the normalization $F_0$ (either the flux at $t_b$ for a broken power-law, or
the normalization at 1-day post-burst) and the power-law indices $\alpha_1$ and  $\alpha_2$.
The sixth column reports the break time t$_{\rm b}$ in seconds since 
the time of trigger (see Table \ref{tab:tess_grb_info}). 
The ``background'' is a nuisance parameter that marginalizes over
residual errors in the background correction. Asterisks indicate light curves with
$F_0$ consistent with 0.}
\centering
\begin{tabular}{cccrccrc}
\hline
\hline
\multicolumn{1}{c}{Identifier} &
\multicolumn{1}{c}{Power-Law} &
\multicolumn{1}{c}{$F_0$} &
\multicolumn{1}{c}{$\alpha_1$} &
\multicolumn{1}{c}{$\alpha_2$} &
\multicolumn{1}{c}{t$_{\rm b}$} &
\multicolumn{1}{c}{Background} &
\multicolumn{1}{c}{$\chi^2$/dof} \\
\multicolumn{1}{c}{} &
\multicolumn{1}{c}{Type} &
\multicolumn{1}{c}{(erg\,cm$^{-2}$\,s$^{-1}$)} &
\multicolumn{1}{c}{} &
\multicolumn{1}{c}{} &
\multicolumn{1}{c}{(s)} &
\multicolumn{1}{c}{(erg\,cm$^{-2}$\,s$^{-1}$)} &
\multicolumn{1}{c}{} \\
\hline
200412B & Broken & $1.18\pm 0.07 \times 10^{-12}$ & $-0.89\pm0.02$ & $-1.79\pm0.15$ & $3.93\pm0.16 \times 10^4$ & $-8.13\pm 3.63 \times 10^{-14}$ & 4.88 \\ 

210204A & Broken & $1.04\pm0.09 \times 10^{-12}$ & $-0.37\pm0.02$ & $-1.18\pm0.15$ & $2.03\pm0.19 \times 10^4$ & $-1.62\pm0.38 \times 10^{-13}$ & 3.38 \\ 

210504A & Broken & $5.16\pm0.53 \times 10^{-13}$ & $0.26\pm0.13$ & $-1.22\pm 0.26$ &  $9.16\pm 1.14 \times 10^3$ & $-6.73 \pm 2.62 \times 10^{-14}$ & 3.23  \\ 

230116D$^*$ & Single & $5.71\pm6.04 \times 10^{-12}$ & $-0.30\pm0.29$ & -- & -- & $-2.69\pm6.40 \times 10^{-13}$ & 15.44 \\ 

230307A & Broken & $1.96\pm0.37 \times 10^{-12}$ & $0.44\pm0.12$ & $-0.12\pm0.03$ & $7.69\pm0.73 \times 10^3$ & $-9.60\pm 3.49 \times 10^{-13}$ & 4.01 \\ 

230903A$^*$ & Single & $1.34\pm2.29 \times 10^{-9}$ & $-1.36\pm0.30$ & -- & -- & $1.68\pm 1.10 \times 10^{-14}$& 2.63\\ 

\hline
\end{tabular}
\label{tab:fit_params}
\end{table*}

\begin{table*}
\caption{Fluxes, fluences, and CRM corrections for the five
GRBs in our sample exhibiting significant deviations from zero flux at the time of trigger that could be 
consistent with prompt emission. The second column is the observed fluence in counts, and
the third column is the total observed burst fluence in TESS (from both prompt and afterglow emission).
The fourth column is the estimated magnitude of the GRB at the time of trigger, calculated
across the entire FFI exposure time. The fifth column is an estimate of the fluence from the afterglow 
in the FFI cadence spanning the time of trigger based on an extrapolation of the best-fit power-law. 
The sixth column gives the corrected flux, by subtracting
column 5 from column 2 and then correcting for the TESS CRM algorithm as described in 
Appendix \ref{app:crm}. The seventh column provides an estimated range for the
magnitude of the prompt emission; this value is calculated using the burst's $T_{\rm 90}$ as 
a lower limit for the emission duration, and
the interval between the trigger time and the end of the contemporaneous FFI as the upper limit.
We note that the
Fermi telemetry limitations for GRB 230307A \citep{dalessi_bad_fermi_times} only allowed us to
estimate a lower limit for the prompt optical flux. GRBs without
a clear afterglow, or those where the afterglow did not contribute significantly to the flux in
the FFI with prompt emission,
are indicated by -- in the relevant columns. We note there may be some contribution from a putative reverse shock
to some of the values for the ``prompt emission'' flux and magnitude given in columns 2 and 4; 
this would overestimate the corrected flux in column 6.}
\centering
\begin{tabular}{cccccrc}
\hline
\hline
\multicolumn{1}{c}{Identifier} &
\multicolumn{1}{c}{Observed Fluence} &
\multicolumn{1}{c}{Total Observed} &
\multicolumn{1}{c}{Peak Observed} &
\multicolumn{1}{c}{Afterglow Contribu-} &
\multicolumn{1}{c}{Corrected} & 
\multicolumn{1}{c}{Corrected $T_{\rm mag}$}
\\
\multicolumn{1}{c}{} &
\multicolumn{1}{c}{in Prompt FFI} &
\multicolumn{1}{c}{Optical Fluence} &
\multicolumn{1}{c}{Magnitude} &
\multicolumn{1}{c}{tion in Prompt FFI} &
\multicolumn{1}{c}{Prompt Flux} &
\multicolumn{1}{c}{of Prompt}
 \\
\multicolumn{1}{c}{} &
\multicolumn{1}{c}{(counts)} &
\multicolumn{1}{c}{(counts)} &
\multicolumn{1}{c}{(TESS band)} &
\multicolumn{1}{c}{(counts)} &
\multicolumn{1}{c}{(counts)} &
\multicolumn{1}{c}{Emission}
\\
\hline
200412B & $2.95 \pm 0.01 \times 10^6$ & $4.6 \pm 0.1 \times 10^6$ & $12.15\pm0.05$ & $5.60\pm0.05 \times 10^5$  & 4.50\,$\pm\,0.10 \times 10^6$ & 5.7--11.5 \\ 

200901A & $3.10\pm 0.51 \times 10^3$ & -- & $18.4\pm0.3$ & - & $9.30\pm0.50 \times 10^3$ & 13.8--15.6 \\ 

210204A & $3.83 \pm 0.07 \times 10^4$ & $6.5 \pm 1.0 \times 10^5$ & $15.67\pm0.05$ & $2.02\pm0.35 \times 10^4$ & 2.13\,$\pm\,0.20 \times 10^4$ & 14.6--16.5 \\ 

220623A & $1.68 \pm 0.10 \times 10^4$ & -- & $16.57\pm0.07$ & -- & $2.28 \pm 0.20 \times 10^4$ & 13.8--15.6 \\ 

230307A & $3.74 \pm 0.03 \times 10^4$ &  $2.9\pm 0.3 \times 10^5$ & $14.51\pm0.05$ & -- & $>$ 4.70\,$\pm\,0.30 \times 10^4$ & 12.6--13.4\\

230903A & $3.03 \pm 0.30 \times 10^3$ & -- & $17.20\pm0.10$ & -- & $3.85 \pm 0.40 \times 10^3$ & 12.5--16.8 \\
\hline
\end{tabular}
\label{tab:flux_ratios}
\end{table*}

For each of the eight GRBs with light curves in Figures \ref{fig:prompt_count_mags} and \ref{fig:afterglow_count_mags}, 
we analyzed the prompt emission and afterglow signatures
in the TESS light curve. We first fit a broken power law to observed afterglows 
\citep{beuermann_1999,li_afterglow_2012}. We also fit single power-laws to these afterglows,
and chose the fit with the lower value of $\chi^2$/dof. We fit
data to the following functional forms, for a broken and
single power law, respectively:
\begin{equation}
    \label{eqn:bpl}
    F(t) = F_0\left[\left(\frac{t}{t_b}\right)^{\alpha_1\omega} + \left(\frac{t}{t_b}\right)^{\alpha_2\omega}\right]+ B,
\end{equation}
and
\begin{equation}
    \label{eqn:spl}
    F(t) = F_0 \left(t^{-\alpha_1}\right) + B.
\end{equation}
Here, $\alpha_1$ and $\alpha_2$ are the power-law indices, $t_b$ is the time of power-law
steepening or turnover, $\omega$ is a smoothing parameter (that was held fixed at 10), $F_0$ is the 
flux normalization at $t_b$, or the amplitude of a single power-law, and $B$ is a nuisance
parameter to marginalize over residuals in the background correction.

Table \ref{tab:fit_params} reports the best-fit parameters for the GRBs
where an afterglow could be modeled by either a single or a broken power-law;
in the case of GRB 200412B, we excluded all points that were visibly part of a flare.
None of these power-law slopes are consistent with zero at the 1-$\sigma$ level.
As part of our analysis of each GRB, we also characterized the effects of the TESS CRM
strategy on the observed prompt optical flux, and established constraints on the
``total'' optical fluence (in counts) that would have been observed from each GRB had the CRM 
clipping not occurred. The results of our analysis of the prompt emission
are reported in Table \ref{tab:flux_ratios}.

\subsection{GRB 200412B}

The Fermi Gamma-Ray Space Telescope triggered on this long GRB at 09:08:40 UT on 12 Apr 2020 
(BTJD 1951.88161; \citealt{fermi_gcn}). The fluence (10--1000 keV) was 8.00\,$\pm$\,0.04 
$\times\,10^{-5}$ erg\,cm$^{-2}$ \citep{fermi_gbm_properties_gcn}. 
An afterglow was detected by Swift-XRT \citep{swift_xrt_afterglow};
this afterglow fell within TESS's field of view
for Sector 23, as well as the Northern Continuous Viewing Zone---it was observed
from Sectors 14--26.
Significant ground-based follow-up was performed by multiple observatories
after the initial discovery of an optical transient by \citet{200412b_master,200412b_master_amur};
these observations were reported by
\citet{200412b_tautenburg_gcn,mondy_terskol_abao_200412b,200412b_late_time_gcn,
mitsume_200412b_gcn,moskvitin_200412b,xin_200412b}, among others.
The measurements from the Devasthal Optical Telescope were published in
Table 1 of \citet{200412b_late_time_optical}.

\begin{figure*}
    \centering
    \includegraphics[width=\textwidth]{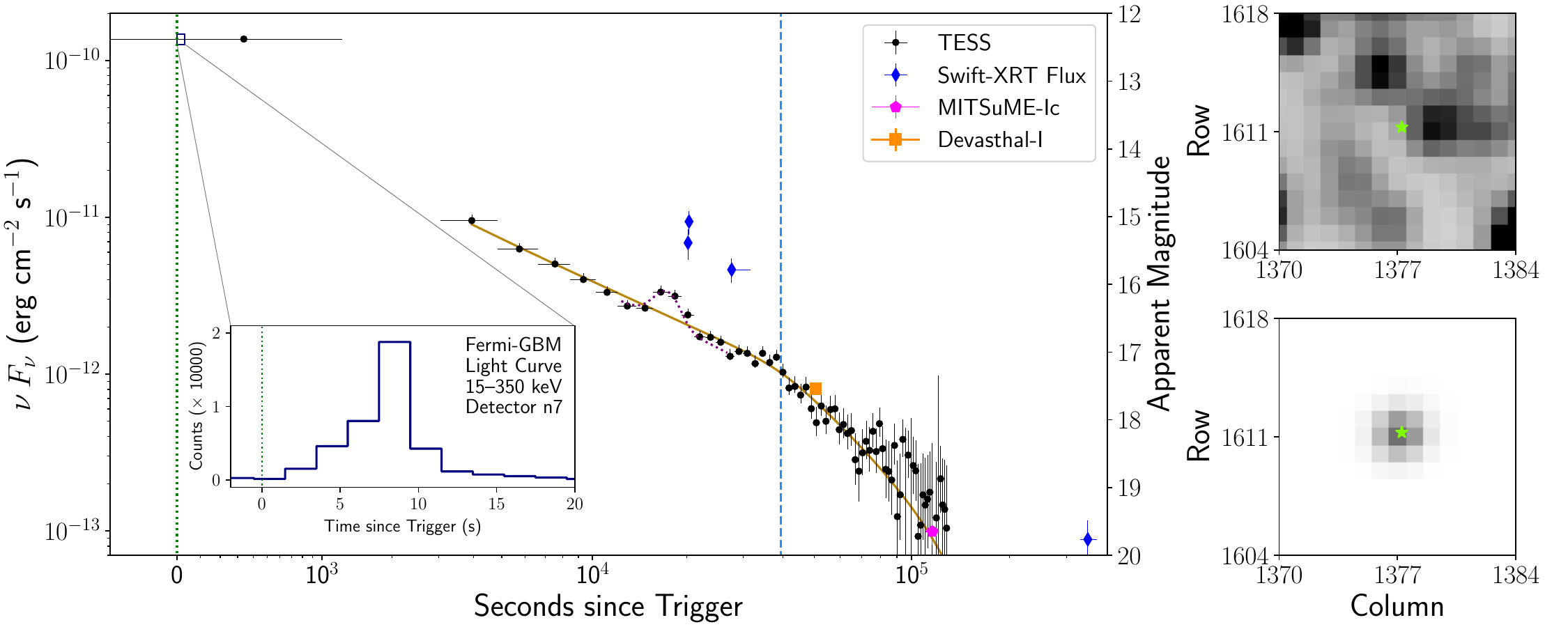}
    \caption{\textbf{Left}: The flux-calibrated, extinction-corrected TESS and Swift-XRT light curves of GRB 200412B. 
    The x-axis measures the seconds since the time of Fermi-GBM trigger, indicated by a green dotted line.
    The best-fit broken power law for the afterglow is shown by the solid gold line. We also show a 
    Gaussian fit to the flare with a purple dotted line. The vertical blue dashed line shows the 
    break time $t_b$ of the broken power law fit. The Fermi-GBM light curve is shown in 
    the inset, and the small blue rectangle shows the 20\,s window expanded in the inset. We also include 
    photometric measurements in the Cousins $I$ band (\citealt{dfot_200412b_gcn}, orange square; 
    \citealt{mitsume_200412b_gcn}, pink pentagon). \textbf{Right}: The top
    panel shows the reference image for the GRB location (from Swift-XRT) in TESS Sector 23, while 
    the bottom panel shows the difference image from the 
    first point in the light curve; there is a clear point source at that position in the difference 
    image. The green star indicates the location of the GRB from Swift-XRT. 
    Reference images were created for each CCD in a sector by median-stacking 
    20 FFIs with low backgrounds from throughout that sector \citep{fausnaugh_diff_imaging}.}
    \label{fig:grb_200412b}
\end{figure*} 

\subsubsection{The light curve}

The emission from this GRB in TESS's band peaked at an apparent magnitude of 12.15$\pm$0.05, 
and decayed over the following day (Fig. \ref{fig:grb_200412b}); this burst also exhibited
a flare between 1.5\,and\,2.1$\times\,10^4$\,s post-trigger.
The peak extinction-corrected flux was 1.47 $\pm$ 0.07 $\times 10^{-10}$
erg\,cm$^{-2}$\,s$^{-1}$. This is approximately 50\% lower than the peak estimate
reported in \citet{roxburgh_afterglows}, who used {\tt tess-reduce} \citep{tessreduce}
for their analysis. Part of this difference may be explained by the zeropoint calculated
by {\tt tess-reduce}, which is 20.31 mag, instead of the value of 20.44 from the
Instrument Handbook. A parallel analysis that we performed 
with {\tt tess-reduce} yields a peak flux that is $\sim$20\% greater than
our calculated value, which corresponds to a magnitude difference of 0.18. 
At this time, it is unclear why there exists a difference
between the {\tt tess-reduce} results and our values; a more detailed comparison of
{\tt tess-reduce} and our pipeline is given in Appendix A.2 of \citet{fausnaugh_ia_2023}.
In the interest of completeness, we also present (extinction-uncorrected) $i$-band measurements in
Figure \ref{fig:grb_200412b} that were reported by \citet{mitsume_200412b_gcn} and
\citet{dfot_200412b_gcn}.

To fit the observed flare, we subtracted the best-fit broken power-law for the afterglow 
from the data and fit a Gaussian, with a constant offset term, to the points corresponding to the flare.
The best-fit parameters for the broken power-law fit to the TESS light curve are given in
Table \ref{tab:fit_params}, and parameters for the fit to the flare (peaking at $\sim$17\,ks
post-burst) are given in Table \ref{tab:flare_fit}.
The high $\chi^2$/dof for the broken power law likely arises from outliers, such as the points near the
power-law break, at $\sim$\,4\,$\times10^4$\,s
post-burst. Both fits---to the afterglow and the flare---are shown in Figure \ref{fig:grb_200412b}.
The fits reported in Table \ref{tab:fit_params} include all points except the flux in the FFI from 
the time at trigger, which deviates from the best-fit power law by over 7-$\sigma$. 
Excluding the second point after the trigger (which shows a 2-$\sigma$ excess over
the current best-fit broken power law) results in a marginally better fit.
Given the timescale on which optical reverse shock
emission occurs---hundreds of seconds (see, e.g., Fig. 1 in \citealt{sari_piran_990123}, and Fig. 
3 in \citealt{yi_reverse_shock_sample})---we suggest that a reverse shock could plausibly explain some
fraction of this observed flux excess. Additionally, excluding the second point from the fit
yields $\alpha_2 = 1.79\pm0.15$, which is consistent with measurements at 1.5, 2.5, and 3.5 days
from the Tautenburg Observatory to within 2-$\sigma$. Those observations yielded a decay slope of $\alpha = 1.582\pm0.004$
\citep{200412b_tautenburg_gcn}.

Swift-XRT observations of the afterglow concurrent with the TESS observation are
also shown in Figure \ref{fig:grb_200412b}. 
Physical explanations for the observed break (at $\sim 4\times 10^4$\,s) are discussed 
further in Section \ref{sec:physical_params}.

\begin{table}[]
\caption{Best-fit parameters, as in Table \ref{tab:fit_params}, for a {Gaussian} 
fit to the flare in GRB 200412B. This fit has a $\chi^2$/dof of 8.7, likely due to
the small number of points that were used.}
    \centering
\label{tab:flare_fit}
\begin{tabular}{cr@{$\pm$}l}
    \hline
    \hline
    \multicolumn{1}{c}{Parameter} &
    \multicolumn{2}{c}{Value} \\
    \hline 
    $F_0$ (erg\,s$^{-1}$\,cm$^{-2}$) & $1.34$\,&\,$0.09\times10^{-12}$ \\
    Time of peak (s) & $1.79$\,&\,$0.01\times10^4$ \\
    Peak width (s) & $1.82$\,&\,$0.18\times10^3$ \\
    Background (erg\,s$^{-1}$\,cm$^{-2}$) & $-3.02$\,&\,$0.86\times10^{-14}$ \\
    \hline 
    \end{tabular}
\end{table} 


\subsubsection{The prompt emission}
\label{subsubsec:200412b_prompt}
The GRB trigger occurred 561.51\,s after the FFI start, and the gamma-ray emission only lasted for a total of $\lesssim$20\,s (as seen in the inset in Figure \ref{fig:grb_200412b}). Thus, the observed flux in the 30\,min FFI that includes the time of trigger contains
contributions from emission occurring on shorter timescales (internal or reverse
shocks), as well as the afterglow. 
To quantify this effect, we extrapolate the afterglow 
power-law to the approximate end of the prompt gamma-ray emission (600\,s after the start of the FFI, or 
$\sim$39\,s after the trigger time) and calculate the fluence of the afterglow during this time, in counts;
these values are listed in Table \ref{tab:flux_ratios}. We find that the afterglow contributes 
approximately $5.60\times 10^5$ counts in this FFI ($\sim$\,20\%).
The remainder is shorter-timescale emission.

As described in Appendix \ref{app:crm}, we correct for the CRM within the two 20-second 
clipping intervals that span the entirety 
of the high-energy emission. We find that the CRM 
clips roughly 40\% of the flux that would have been recorded as part of the prompt emission, 
as shown in Figure \ref{fig:crm_clipping_example}). 
Correcting for the clipped flux, we can constrain a range of values for the true magnitude 
of the prompt emission based on the emission duration. The lower limit for the duration of optical
emission can be 
approximated by the T$_{\rm 90}$ value, and the upper limit is the difference between the end
of the FFI and the trigger time. This spans 3 orders of magnitude for this burst
(6 to $\sim\,1250$\,s), so the magnitude is highly uncertain. We find that the
prompt flash could have had a magnitude of anywhere from 5.7 to 11.5. In the 2-second exposure 
with the highest flux (which was clipped by the CRM), the instantaneous 
magnitude could have been as bright as $5.5$ assuming that there is no reverse shock emission 
component in this FFI.
However, if the shape of the prompt optical light curve deviates from that of the high-
energy light curve, or the flux excess (partially) arises from a reverse shock, our flux 
corrections are systematically overestimated, and the magnitude of the prompt emission would 
actually be fainter than these estimates. The true value would then
depend on the number of emission components and their relative strengths.




\subsection{GRB 200901A}

\begin{figure*}
    \centering
    \includegraphics[width=\textwidth]{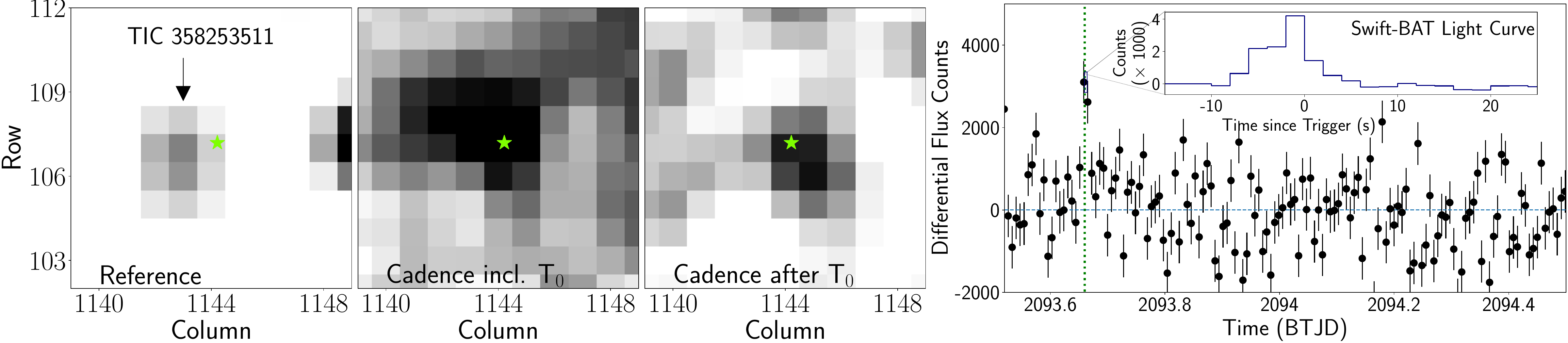}
    \caption{TESS images and light curve for GRB 200901A. The leftmost panel
    is the reference image at the location of the Swift-XRT localized 
    afterglow; the next panel to the right is the difference image during
    the time of trigger, and the panel to the right of that is the difference
    image after the time of trigger, which shows some evidence for a source 
    at that position. The light profile of the source at that location 
    is distorted compared to the TESS pixel response function.
    The rightmost panel shows the TESS light curve, with the Swift-BAT light curve
    as an inset, and the trigger time marked by a dotted green line. The dashed
    blue line denotes zero flux. A potential contaminating star, discussed in 
    Section \ref{subsec:identifying_signals}, is labeled in the reference image.}
    \label{fig:grb_200901a}
\end{figure*}

Swift-BAT triggered on this long GRB at 03:47:31 UT on 1 September 2020 (BTJD 2093.65995; 
\citealt{swift_200901a}). Using the Fermi-GBM data, \citet{fermi_200901a} reported a fluence 
in the 10--1000 keV band of 9.4\,$\pm$\,0.54\,$\times$\,10$^{-6}$\,erg\,cm$^{-2}$. 
The MASTER telescope network reported the detection of an optical transient 
at the Swift-XRT location 20 minutes after the trigger 
\citep{master_200901a}, and Swift-UVOT reported a faint transient
50\,min after the trigger \citep{200901a_uvot}.

The location of this GRB fell in TESS's field of view for Sector 29, and the light curve
around the time of trigger is shown in Figure \ref{fig:prompt_count_mags}(b), and 
Figure \ref{fig:grb_200901a}.
The trigger occurred $\sim$\,140\,s prior to the end of the TESS FFI. 
In this FFI, the optical magnitude is $T=18.4\pm0.3$; this detection is 3.8-$\sigma$. 
There is also a 3.3-$\sigma$ flux excess 
in the subsequent FFI ($T=18.6\pm0.3$). These magnitudes are consistent with the limit of $R\sim17.8$ at 21.3\,min
post-burst from \citet{200901a_fram-auger_gcn}. Inspection of the difference image
reveals a source at that location, although the light profile is somewhat distorted compared
to the TESS pixel response function (PRF), as visible in Figure \ref{fig:grb_200901a}.

We used the Swift-BAT light curve to evaluate the effects of CRM on this burst.
The trigger happened roughly in the middle of the emission, as visible in the inset of
Figure \ref{fig:grb_200901a}---almost 102\,s
before the end of the FFI. The emission started approximately 110\,s before the end of
the FFI. We find that the CRM causes the fluence to be underestimated
by around a factor of 3. The brightest 2-s optical flash might have had a magnitude of
$\sim$\,11.8. Our estimated range for the corrected magnitude based on the $T_{\rm 90}$ and
the interval between the start of emission and the end of the FFI is 13.8--15.6.

\subsection{GRB 210204A/AT2021buv}

\begin{figure*}
    \centering
    \includegraphics[width=\textwidth]{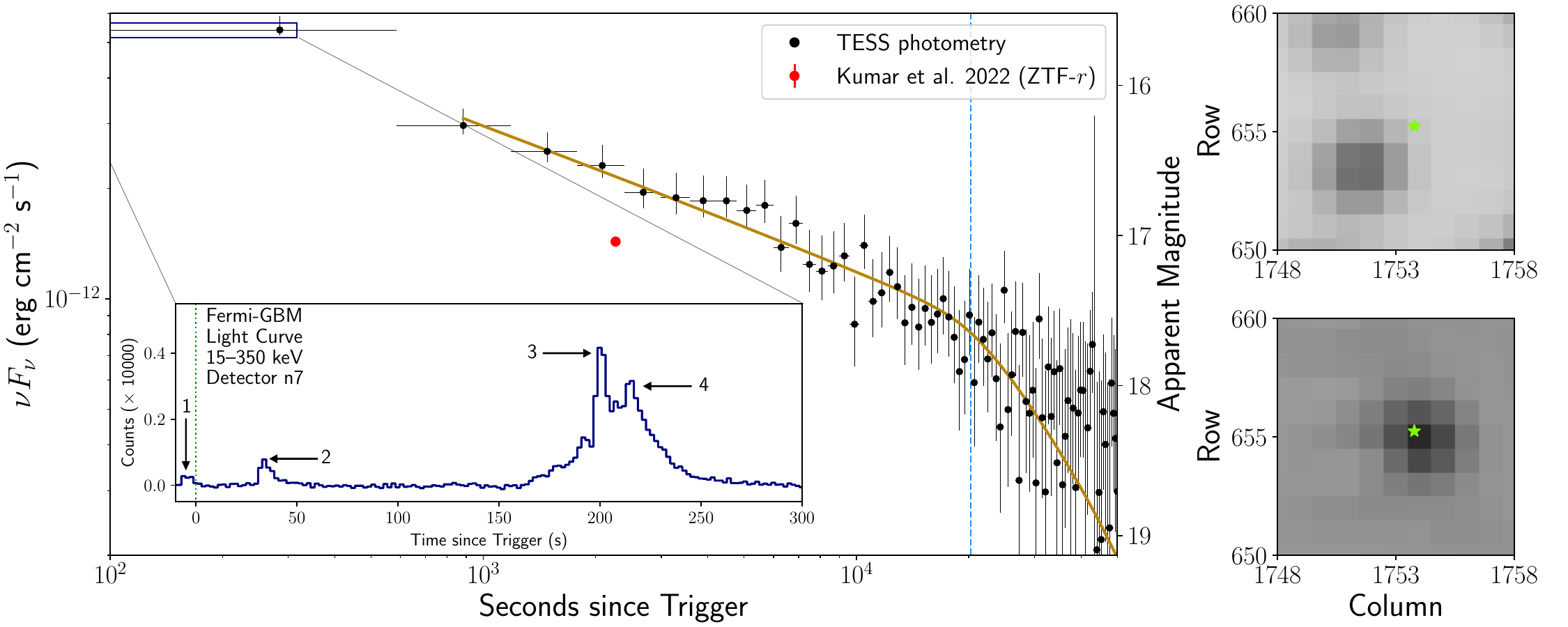}
    \caption{Same as Figure \ref{fig:grb_200412b}, but for GRB 210204A, including
    an extinction-corrected $r$-band afterglow measurement from \citet{kumar_afterglow_flares} 
    for comparison purposes (red dot). The four distinct peaks visible in the high-energy
    light curve are marked and numbered in the inset. The Swift-XRT light curve starts $10^5$\,s after
    the trigger time, at which point the afterglow had faded beyond the TESS detection limit, so
    it is not shown here.
    We find evidence for a temporal break at $\sim$21\,ks post-burst, 
    but these points are close to TESS's detection limit, so the steepened power-law
    index is more uncertain. We do not show further observations from \citeauthor{kumar_afterglow_flares},
    who analyze data out to 10$^6$\,s post-burst; these data points would significantly affect the scale 
    of this plot, so we direct the reader to their Figure 6 for the full light curve.}
    \label{fig:grb_210204a}
\end{figure*}

Fermi-GBM initially triggered on this long GRB at 06:29:25 UT on 4 Feb 2021 (BTJD 
2249.77660; \citealt{fermi_210204_gcn}). The fluence was 
$5.76\pm0.04\times10^{-6}$\,erg\,cm$^{-2}$ in the 10--1000 keV band;
its T$_{90}$ of 206\,s makes it the longest GRB in our sample with a 
confident detection in TESS. Forty-five minutes after the GRB trigger, ZTF independently 
triggered on an optical transient with 
magnitude $r$=17.11, located within the Fermi localization region, as part of the
Caltech-TESS shadowing survey \citep{caltech_tess_shadowing}.
Swift also detected an uncatalogued X-ray source at the ZTF transient location \citep{kennea_210204a},
further strengthening the association between this transient and the Fermi-GBM GRB. 
\citet{xu_redshift_210204a_association} reported a redshift of $z = 0.876$ using X-shooter at the
Very Large Telescope (VLT).
The results of significant multi-wavelength follow-up are presented in \citet{kumar_afterglow_flares},
and light curves from radio to X-rays are shown in Figure 6 of their paper. 
The Fermi-GBM light curve is shown in the inset of our Figure \ref{fig:grb_210204a}, and the ZTF light 
curve of the afterglow is given in Figure 4 of \citet{andreoni_ztfrest}.

After performing the barycentric correction, we found that the prompt emission for
this GRB lies wholly within a single 10\,min FFI cadence. The high-energy light curve exhibits 
4 distinct peaks (labeled in the inset of Figure \ref{fig:grb_210204a}); 
Fermi triggered after the first. The gamma-ray peak with the highest fluence
starts $\sim$160\,s after the trigger.
The FFI cadence with peak flux has a brightness of T$_{\rm mag}$ = 15.67\,$\pm$\,0.05,
without accounting for CRM.

As with GRB 200412B, we fit a broken power-law to the light curve, both with and 
without the initial point (with maximum flux). The fits that
include the first point are somewhat worse than those without
this point, with $\chi^2/{\rm dof}\,\sim\,3.74$. There may be comparable contributions
from both the afterglow and prompt emission in this cadence. Additionally, 
there may be a plateau in the TESS light curve from $3-6\times10^3$\,s
post-trigger. Our power-law fits do not substantially change when including
or excluding this feature, which suggests that this feature is statistically
insignificant.

The best fit to the TESS data is shown in Figure
\ref{fig:grb_210204a}, and the corresponding parameters are given in
Table \ref{tab:fit_params}. We also jointly fit the TESS data and the
$i$-band data from the observing log in Table A.1 of \citeauthor{kumar_afterglow_flares} This
yields a revised estimate of $\alpha_2 = -0.76\pm0.04$, which is inconsistent with 
their results for the late-time power-law index. The time of the power-law break also shifted 
and became more uncertain, to $t_b = 8.7\pm2.1 \times 10^3$\,s post-burst. While 
$t_b\,\sim\,0.1$\,d is consistent with the weak constraint 
from \citet{kumar_afterglow_flares} ($t_b = 0.37\pm0.3$\,d), the late-time
decay index is not.

Given the presence of both a break in the TESS data, as well as another break observed by
\citeauthor{kumar_afterglow_flares}, we also evaluated whether a triple power-law fit 
to the data (Eqns. 2--4 in \citealt{li_afterglow_2012}) 
was a better match to the observations. We find through the use of an $f$-test that this
fit is marginally favored (at a $\sim$70\% confidence level). The two breaks occur at
$8.5\pm2.7\times10^3$\,s post-burst and $1.87\pm0.54\times10^5$\,s post-burst; the latter
break occurs between the end of the TESS observations and the start of the 
\citeauthor{kumar_afterglow_flares} observations. The three power-law indices are
$\alpha_1 = -0.33\pm0.05$, $\alpha_2 = -0.78\pm0.06$, and $\alpha_3 = -1.25\pm0.06$. 
The value for $\alpha_3$ is consistent with the decay index found by 
\citeauthor{kumar_afterglow_flares} for the data at that time.
Such a three-component power-law could possibly be 
explained by early-time energy injection, followed by a ``normal'' decay, and 
then a subsequent jet break \citep{zhang_afterglow}.

After removing contributions from the afterglow fluence in the TESS FFI (the afterglow
was assumed to start at $\sim300$\,s post-burst, which marked the end of the gamma-ray emission
in the Fermi-GBM light curve), we found
that the CRM removes 15\% of the prompt optical flux, under the assumption that it has
the same shape as the high-energy light curve. We also calculated the magnitudes of the brightest 2-s 
optical flashes that would accompany each of the four peaks seen in the prompt emission, 
and found values of $\sim$16.15, 15.02, 13.20, and 13.50. After these corrections,
the magnitude of the optical prompt emission detected during the FFI was calculated
to be 15.8, when integrated over the duration of gamma-ray emission ($\sim$\,300\,s).
However, since the duration of the prompt optical emission is uncertain, and could
range from 100\,s to 600\,s (the duration of the brightest peak to the duration of the FFI),
this magnitude could range between 14.6 and 16.5.

Physical interpretations of the break seen in the TESS
light curve are discussed in Section \ref{subsec:210204a_physical}, as are the
discrepancies between the TESS light curve and the predictions for the early-time
light curve from \citeauthor{kumar_afterglow_flares}

\subsection{GRB 210504A}
\label{subsec:210504a}

Swift-BAT detected the long GRB 210504A at 13:54:53 UTC on 2021 May 4 (BTJD 2339.08622; 
\citealt{210504a_bat}); the X-ray afterglow was localized with the XRT, and fell 
within the TESS field of view for Sector 38. The fluence in the 15--150 keV 
band was $2.7\pm0.2\times10^{-6}$\,erg\,cm$^{-2}$ \citep{swift_bat_210504a}. 
The Nordic Optical Telescope detected an afterglow 9.3\,h post-burst
\citep{210504a_not_afterglow}, and \citet{210504a_redshift} found
$z = 2.077$ using X-shooter.

\begin{figure}
    \centering
    \includegraphics[width=\linewidth]{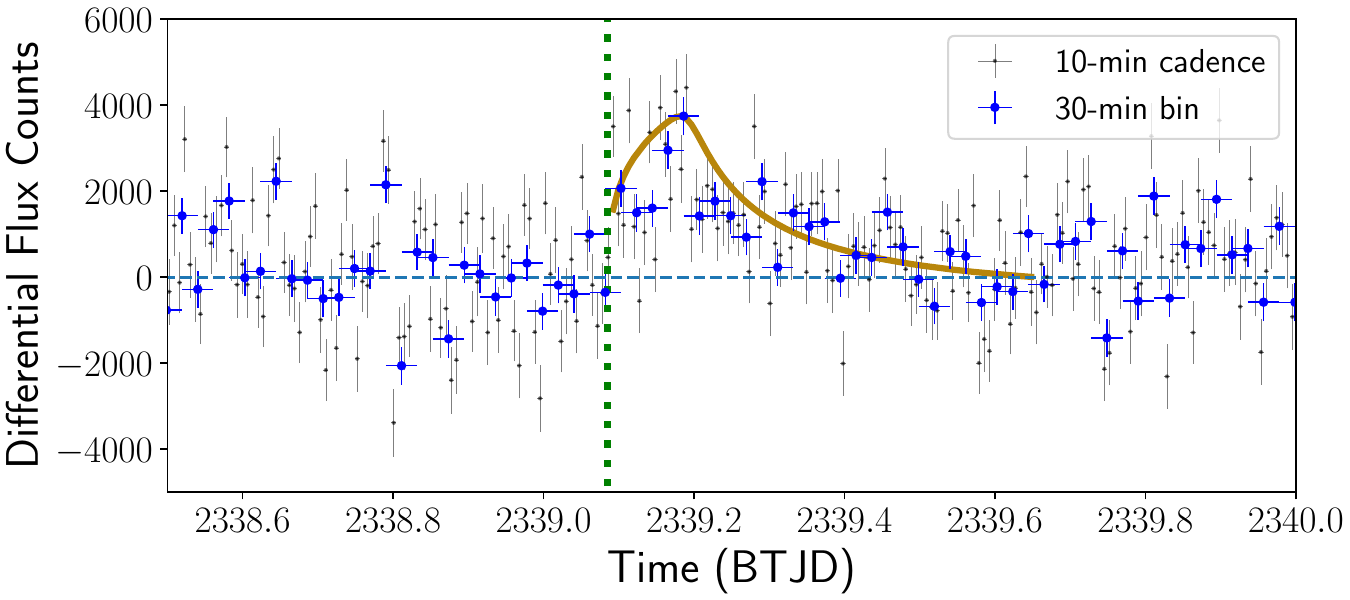}
    \caption{The light curve of GRB 210504A. The light gray points are
    the original light curve at 600-s cadence, and the blue points are the light 
    curve binned to 1800\,s. We also fit a broken power law to the 10\,min
    data and show the resulting fit in gold. This power law peaks at 
    $\sim\,$9\,ks post-burst.}
    \label{fig:grb_210504a}
\end{figure}

Visual inspection of the 10\,min cadence TESS light curve around the time of 
trigger revealed several points above zero flux that were consistent
with the temporal profile of an afterglow rise and decay. 
There are formal 3-$\sigma$ detections of flux excesses in individual cadences 
approximately 2\,hr after the trigger, which correspond to the cadences 
surrounding our best-fit afterglow
peak time. There is also marginal evidence for a point source at the location of the burst 
in the FFI cadences after the trigger.
The temporal coincidence between these measurements and the GRB trigger 
are consistent with an afterglow that peaks $\sim\,9$\,ks after the burst.
The best-fit parameters for the broken power-law are enumerated in Table \ref{tab:fit_params}.
Figure \ref{fig:grb_210504a}
shows both the original light curve and the light curve binned
to 1800\,s, and the best-fit broken power law for the afterglow.

\subsection{GRB 220623A}

\begin{figure*}
    \centering
    \includegraphics[width=\textwidth]{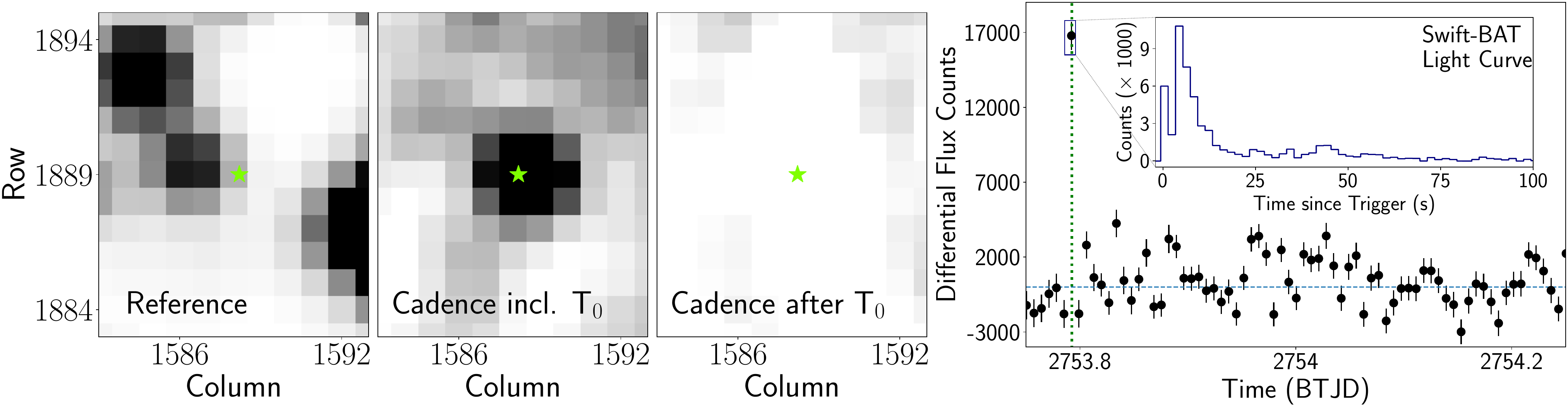}
    \caption{Same as Figure \ref{fig:grb_200901a}, but for GRB 230903A. There
    is a clear source in the difference image at the time of trigger, followed
    by a subsequent non-detection. The Swift-BAT light curve (inset) shows two
    distinct peaks, one of which likely is a precursor.}
    \label{fig:grb_220623a}
\end{figure*}

Swift-BAT triggered on this long-duration burst at 07:04:14 UT on 23 June 2022 (BTJD 2753.79238;
\citealt{220623a_initial_gcn}). The fluence was
$3.5\pm0.1\times10^{-6}$\,erg\,cm$^{-2}$ in the 15--150 keV band \citep{220623a_refined}. 
An optical counterpart detected 47\,s after
the trigger was reported by \citet{220623a_afterglow_gcn}; this was detected in a 1\,s exposure from the 
BOOTES-5/JGT telescope. An afterglow was detected in multiple 1\,s exposures taken between
the trigger and 30\,min post-burst, using the same telescope
(Y.-D. Hu, private communication). The localized burst 
position fell within TESS's field of view for Sector 53.

We detected a 12.9-$\sigma$ deviation from zero flux in the 10\,min TESS 
cadence that spanned the time of the trigger
(rightmost panel of Figure \ref{fig:grb_220623a}). Inspection of the difference images around the 
time of the trigger (second panel from left in Figure \ref{fig:grb_220623a}) reveals a point source 
in the FFI corresponding to the trigger time; this point source vanishes in the subsequent 
FFI. The flux in the FFI corresponds to an extinction-corrected 
T$_{\rm mag}$ =  16.57\,$\pm$\,0.07. While these observations provide clear evidence for 
prompt optical emission, we do not find any indication of an 
accompanying afterglow. The lack of afterglow detection could be caused by the 
sky backgrounds in the FFIs during the 1.2\,d before and after the trigger being 
large due to Earthshine, and the afterglow likely being fainter than the 3-$\sigma$
detection limit of 18.16 in that sector.

After accounting for the effects of CRM, we find that it removes approximately 
35\% of the total flux. The gamma-ray emission has at least two distinct peaks (inset in 
the right panel of Figure \ref{fig:grb_220623a}), with other low-significance peaks starting 24\,s
after the trigger. The brightest 2-s optical flash has a magnitude of $\sim$\,12.2.
Given the uncertainty in
the optical emission duration (with a lower limit being the T$_{\rm 90}$ of 50\,s and an upper limit of
280\,s, the interval between the trigger and the end of the exposure), our estimate for
the true magnitude across the entirety of the prompt flash ranges from 13.8--15.6.

\subsection{GRB 230116D}

\begin{figure}
    \centering
    \includegraphics[width=\linewidth]{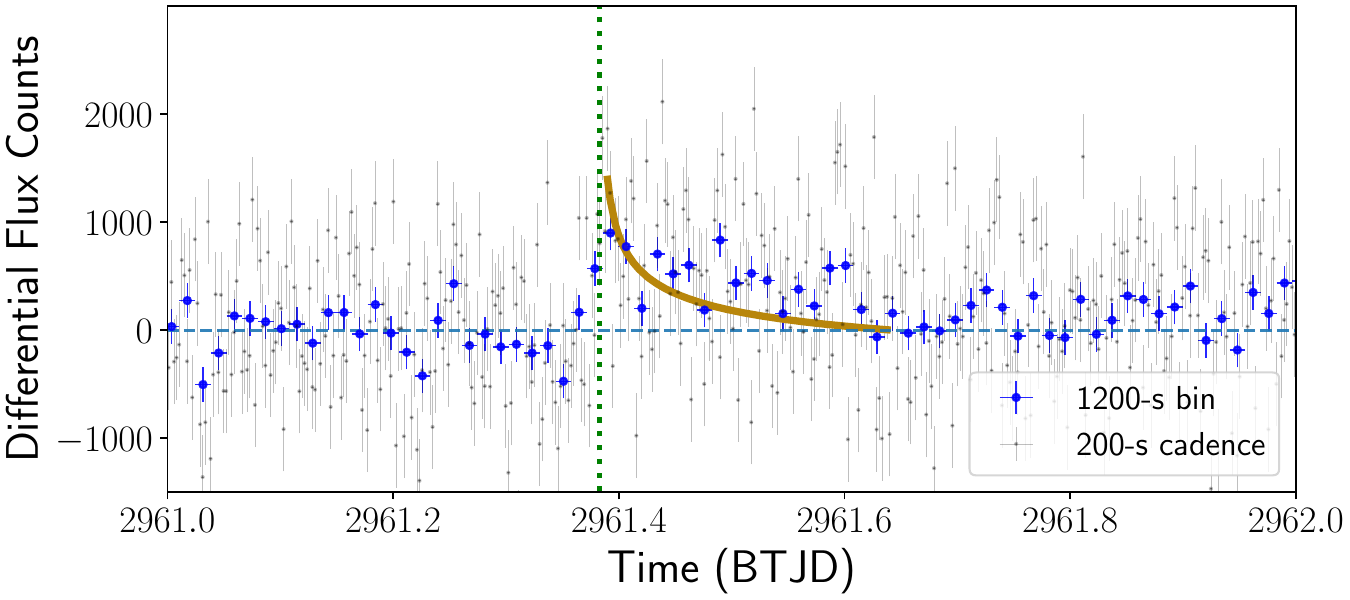}
    \caption{The light curve of GRB 230116D, as in Figure \ref{fig:grb_210504a}; here,
    the blue points are binned to 1200\,s. The fit was performed with the original
    200\,s cadence data.}
    \label{fig:grb_230116d_lc}
\end{figure}

Swift-BAT triggered on this long burst at 21:04:43 UT on 16 January 2023 (BTJD 2961.38386;
\citealt{gcn_230116d}). The measured fluence was $8.12\pm1.2\times10^{-7}$\,
erg\,cm$^{-2}$\,s$^{-1}$ in the 15--150 keV band \citep{swift_bat_230116d}.
\citet{230116D_redshift_gcn} found the burst's redshift to be $z = 3.81$
using the SCORPIO-2 spectrograph on the SAO RAS telescope.
The localized burst position fell within TESS's 
field of view for Sector 60. We
find a 3-$\sigma$ deviation from zero flux around the time of trigger in the TESS data.

The peak flux in the 200-s cadence light curve corresponds to an extinction-corrected
apparent magnitude of 17.71$\pm$0.35, which is slightly above the 3-$\sigma$ detection limit 
of 17.84 given in Table \ref{tab:tess_grb_info}. We fit the 
differential light curve to a single power law (shown in Figure \ref{fig:grb_230116d_lc}, 
with the corresponding parameters in Table \ref{tab:fit_params}). 
The high $\chi^2$/dof could be explained
by the outliers present during the afterglow decay portion of the light curve. Moreover, 
the best-fit $F_0$ is consistent with 0.
A statistical $f$-test, however, provides evidence (at a 95\% confidence level) that 
the single power-law fit is favored over a fit to a constant flux. 
Given that there are three stars nearby with G$_{\rm RP} \sim 18$, noise from 
these stars could also induce significant scatter in the light curve before the GRB,
which has the effect of making
the detection limit brighter.

To verify our results from TESS data, we flux-calibrated the early-time
$r$-band observation from \citet{230116d_early_time}, and found it to
be consistent with the flux-calibrated TESS measurement. Additionally,
we used the $Rc$-band photometry from 1--3 hours post-burst 
reported in \citet{230116D_late_time} 
to verify the power-law index and some of the late-time detections visible in Figure 
\ref{fig:afterglow_count_mags}(d)(ii). Combining the observation of 
\citeauthor{230116d_early_time} and the first observation of 
\citeauthor{230116D_late_time}, we find a decay slope of approximately $-0.28$, 
consistent with our result in Table \ref{tab:fit_params}. The $Rc$-band
observations also show flaring behavior, with two distinct peaks; 
the peak of the first flare could explain one of the confident detections in Figure
\ref{fig:afterglow_count_mags}(d)(ii).

\subsection{GRB 230307A}
\label{subsec:230307a}

\begin{figure*}
    \centering
    \includegraphics[width=\textwidth]{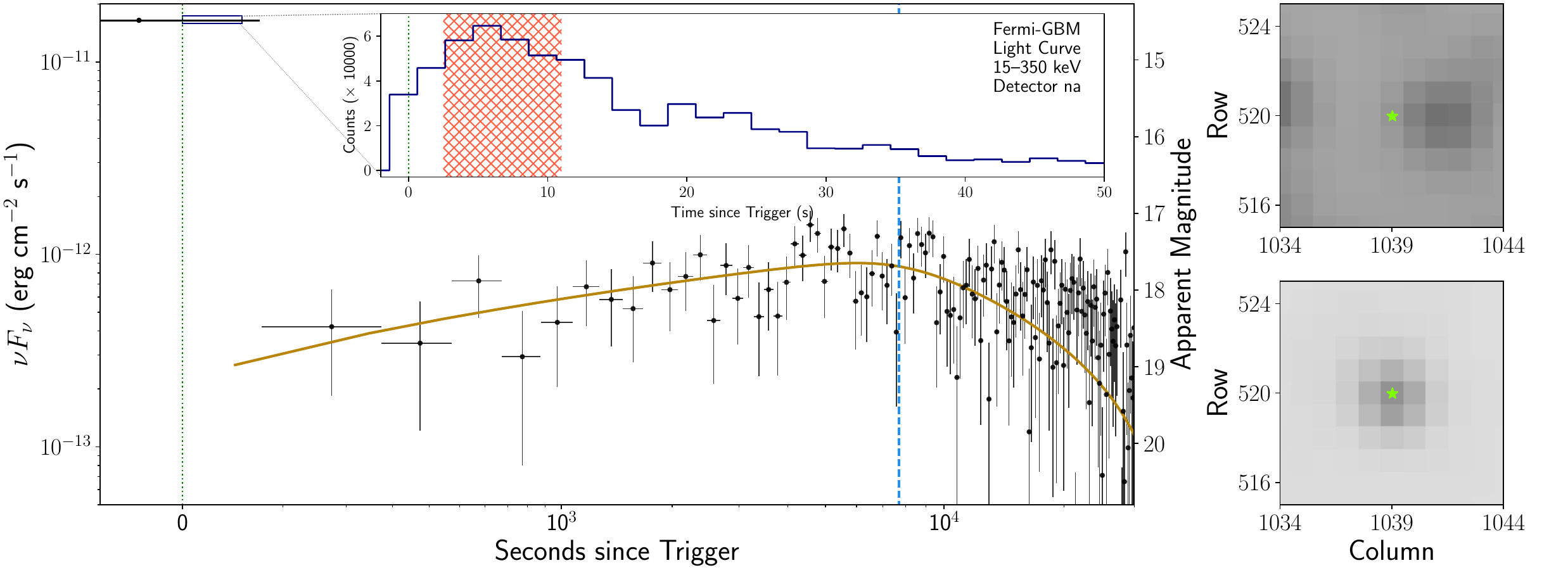}
    \caption{Same as Figures \ref{fig:grb_200412b} and \ref{fig:grb_210204a}, but for GRB 230307A.
    In this case, the blue dashed line represents the peak time of the afterglow fitted by the
    break time parameter $t_b$ in the broken power-law model. The Fermi-GBM TTE
    data from 2.5--11\,s after the trigger are unreliable due to a pulse pile-up issue, as
    described in \citet{dalessi_bad_fermi_times}; these times are marked by red cross-hatching
    in the inset Fermi-GBM light curve.}
    \label{fig:grb_230307a_lc}
\end{figure*}

Fermi-GBM triggered on this long GRB at 15:44:06 UT on 7 Mar 2023 (BTJD 3011.15549; 
\citealt{fermi_gcn_230307a}). The measured fluence was $2.951\pm0.004\times10^{-3}$\,erg\,cm$^{-2}$
in the 10--1000\,keV energy interval \citep{dalessi_fluence}, which makes it
one of the brightest bursts detected. An afterglow was
clearly detected by Swift-XRT \citep{230307a_xrt_afterglow}, and the localized 
afterglow fell within TESS's field of view for Sector 62. The discovery of prompt optical
emission with TESS and an accompanying afterglow was 
reported by \citet{230307a_tess_gcn} via GCN, and
published by \citet{fausnaugh_230307a} and \citet{levan_230307a}.

The TESS light curve displays a prompt emission component, followed by a rise and decay 
that likely correspond to an afterglow. Our best-fit values for the broken power-law fit
to the afterglow are reported in Table \ref{tab:fit_params};
this is plotted in Figure \ref{fig:grb_230307a_lc}.
The prompt optical emission has a magnitude of $T=14.51\pm0.05$. 

For GRB 230307A, the prompt flash in the TESS light curve clearly precedes a 
distinct afterglow signal, so it seems reasonable to assume that the afterglow 
itself contributes negligible flux during the first 200s TESS FFI cadence. 
Extrapolating the afterglow shows that its contribution
accounts for $\lesssim1\%$ of the flux in the first FFI. Thus, we do not subtract 
any afterglow contribution when correcting for CRM. Because the Fermi GBM TTE data from 2.5--11\,s 
after the burst are unreliable due to limits on instrument telemetry \citep{dalessi_bad_fermi_times},
our estimate for the corrected optical flux is only a lower limit.
We find that the CRM algorithm clips $\sim$\,20\% of the counts, which
is almost exactly canceled by the 20\% of exposure time removed and thus
has no net effect on the reported magnitude.

We calculate the limits on the magnitude based
on the T$_{\rm 90}$ (35\,s) and the interval between the trigger and the end of the FFI (72\,s). We find a corrected magnitude range of 12.6--13.4 using the
estimated prompt flux counts, after correcting for CRM.
The brightest 2-s portion of the optical light curve had a 
magnitude of $\sim$\,12. Because the estimates for these magnitudes in the optical
light curve were calculated based on the unreliable TTE data, these estimates are a lower limit on any correlated optical prompt emission and are likely fainter
than the true magnitudes.


\subsection{GRB 230903A}

\begin{figure*}
    \centering
    \includegraphics[width=\textwidth]{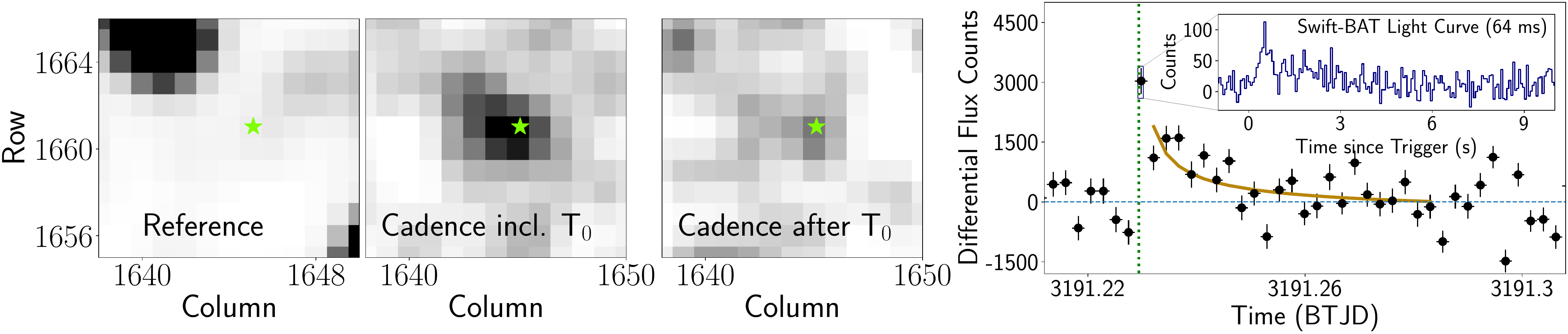}
    \caption{Same as Figure \ref{fig:grb_200901a}, but for GRB 230903A. The light profile of 
    the GRB in the difference image has unusual structure that does not match the
    TESS PRF, but the temporal coincidence suggests that this is a counterpart to the GRB. 
    The Swift-BAT light curve is shown at its 
    native cadence of 64\,ms to highlight the relatively short nature of the burst,
    compared to the other bursts in our sample. The best-fit single power-law model for
    the afterglow (parameters enumerated in Table \ref{tab:fit_params}) is shown in gold.}
    \label{fig:230903a_lc}
\end{figure*}

Both Fermi-GBM and Swift-BAT triggered on this intermediate-duration GRB ($T_{\rm 90}\sim2.5$\,s), 
with Fermi detecting the GRB at 
17:22:58 UT \citep{230903a_fermi_gbm_detection}, and Swift detecting this burst 2 
seconds earlier \citep{230903a_swift_detection}. The time of the Swift-BAT trigger corresponds 
to BTJD 3191.22939. The fluence in the 15--150 keV band
was 2.2$\pm$0.4$\times$10$^{-7}$\,erg\,cm$^{-2}$ \citep{230903a_swift_bat_analysis}.
This burst's spectrum suggests that it is an X-ray rich burst, meaning that the ratio of 
the fluence in the 25--50 keV band to that in the 50--100 keV band is 
between 0.72 and 1.32 \citep{sakamoto_xrr}.
The localized burst position fell within the TESS field of view for Sector 69, 
and the discovery of prompt optical emission at this GRB's position was reported
by \citet{230903A_tess_gcn}.

We find that the prompt optical emission in the 200-s FFI cadence at the time of burst 
(using the Swift-BAT trigger time)
has an apparent magnitude of $T=17.2\pm0.1$ in 
the TESS band. While this cadence was the only point in the light curve to have a significant
detection above the 3-$\sigma$ level (T$_{\rm mag}$ = 17.83), we still fit a power law to the data from 
$t_0 + 100$\,s (where $t_0$ is the time of trigger) to $t_0 + 5\times10^3$\,s. 
We conducted a statistical $f$-test and found that a single power law was somewhat 
preferred (at a $\sim99$\% confidence level) over a constant line. 
Including the first point (i.e., the prompt detection) in the fit results
in a statistically indistinguishable fit, increasing the $\chi^2$/dof by 0.01.
The best-fit power-law is shown in gold in the rightmost panel of Figure 
\ref{fig:230903a_lc}.

To evaluate the effects of CRM on the observed flux, we used the Swift-BAT light
curve rather than the Fermi-GBM light curve due to (a) its more significant detection
of the GRB, and (b) its finer time resolution. We do not include any afterglow
contribution in the prompt FFI, because the extrapolation (from the model that
excludes the first point) would lead to more flux
than was actually observed at the time of trigger. The Swift-BAT trigger 
occurred 62.85\,s after the start of the FFI,
and there was only one episode of gamma-ray emission lasting approximately 2 seconds.
We find that the TESS flux is underestimated by approximately
27\% due to the effects of CRM. The brightest 2-s optical flash has a magnitude of 
$\sim$\,15.3\footnote{While the T$_{\rm 90}$
of this burst is only $\sim2.5$\,s, the main $\gamma$-ray emission was split over two 2-s TESS
sub-exposures.}. Our estimated range for the true magnitude based on the T$_{\rm 90}$, and the
interval between the trigger and the end of the FFI, is 12.5--16.8. This large 
range arises from the 135\,s uncertainty in the duration of 
optical emission.

\section{Constraints on physical parameters}

\label{sec:physical_params}
The TESS light curves of four GRBs from Section \ref{sec:grb_lcs} provide enough
information for us to constrain physical parameters of those bursts, including
the initial Lorentz factor $\Gamma_0$ and the electron index $p$.
Such an analysis is possible for GRBs 
200412B, 210204A, 210504A, and 230307A. Where possible, we compare our derived
constraints on $\Gamma_0$ and $p$ to those available in the literature. In some cases, we 
utilize fiducial parameter values or values from the literature in our calculations.


\subsection{GRB 200412B}
\label{subsec:200412b_physical_params}
From the light curve of GRB 200412B, we can constrain the initial bulk
Lorentz factor $\Gamma_0$ and evaluate the potential physical mechanism
underlying the observed power-law break in the light curve.

\subsubsection{The Initial Bulk Lorentz Factor $\Gamma_0$}

The time of afterglow peak indicates when the GRB fireball begins to
decelerate, and the initial bulk Lorentz factor $\Gamma_0$ begins to
change. Given that we do not observe a clear peak in the TESS light curve
of this burst's afterglow (cf. GRB 230307A), we can infer 
that $t_{\rm peak} \lesssim 1200$\,s---the amount of time between
the trigger ($t = 0$) and the end of the FFI cadence spanning the initial emission.
The expression for $\Gamma_0$, assuming a constant density circum-burst medium, is given in 
\citet{sari_piran_prompt} and \citet{molinari_lorentz_fireball}:

\begin{align}
\label{eq:lorentz}
\begin{split}
    \Gamma(t_{\rm peak}) &= \left[\frac{3 E_{\gamma, {\rm iso}} (1+z)^3}{32\pi n m_p c^5 \eta t_{\rm peak}^3}\right]^{\frac{1}{8}} \\
    &\approx  160\left[ \frac{E_{\gamma, {\rm iso,53}} (1+z)^3}{\eta_{0.2} n_0 t_{\rm peak, 2}^3}\right]^{\frac{1}{8}}
\end{split}
\end{align}

Here, $E_{\gamma, {\rm iso}} = 10^{53} E_{\gamma, {\rm iso,53}}$ is the isotropic energy released
in gamma rays, $\eta=0.2\eta_{0.2}$ is the radiative efficiency,
$t_{\rm peak} = 100\,t_{\rm peak, 2}$ is the time of afterglow peak (in s), $z$ is the redshift,
$n_0$ is the particle density of the surrounding medium, and $m_p$ is the proton mass.
If we assume a fiducial GRB isotropic energy of 10$^{53}$\,erg, 
$\Gamma_0 = A \times (1+z)^{3/8} (\eta_{\rm 0.2} n_0)^{-1/8}$. Here, we have introduced a
constant $A$ that encapsulates estimates
for $t_{\rm peak}$ ranging from 40\,s post-burst to 1200\,s post-burst (i.e., the end of the FFI cadence),
without making assumptions about $z$, $\eta$, or $n_0$. Depending on the
value of the peak time, $A$ can range from 63 to 225, with a lower value for $A$ corresponding to a later $t_{\rm peak}$.
Identification of a host galaxy and spectroscopic measurements of
its redshift can yield a revised estimate of $\Gamma_0$ that may better reflect the
true value, given its comparatively weak dependence on $\eta_{\rm 0.2}$ and $n_0$.

The alternative model to a homogeneous circum-burst medium is a wind medium, with a 
radial density power-law profile that can be written $n(r) = n_0 r^{-2}$. In this case,
we must use a generalized version of Equation \ref{eq:lorentz} 
\citep{nava_2013, ghirlanda_lorentz_2018} that introduces a power-law index $s$, which
equals 2 in the wind case:

\begin{equation}
    \Gamma_0 = \left[\frac{(17-4s)(9-2s)3^{2-s}}{2^{10-2s} \pi (4-s)} \left(\frac{E_0}{n_0 m_p c^{5-s}}\right)\right]^{\frac{1}{8-2s}} t_{\rm p,z,2}^{-\frac{3-s}{8-2s}}.
\end{equation}
Here, $t_{\rm p, z, 2}$ is the redshift-corrected time normalized to 100\,s ($t_{\rm p, 2}/(1+z)$).
For this case, the dependence on $n_0$ and $z$ can be rewritten with a different 
constant $B$, which takes into account our fiducial value of E$_{\rm iso}$ and our range of $t_{\rm peak}$, and
the normalization $n_0\sim10^{35}$\,cm$^{-1}$ given in Section 6 of \citet{ghirlanda_lorentz_2018}.
We find that $\Gamma_{\rm 0, wind} = B \times (1+z)^{1/4}$, with $B$ ranging from 40--90 for
our assumed values of $t_{\rm p}$. Lorentz factors for bursts with a wind medium 
must be lower than those for bursts occurring in homogeneous mediums.
An empirical distribution of Lorentz factors
for different circum-burst mediums is shown in Figure 6 of \citet{ghirlanda_lorentz_2018}.

\subsubsection{The temporal power-law break}
\label{subsubsec:200412b_temporal_break}

The clear change in the power-law decay slope observed in the light curve of 
GRB 200412B (blue dashed line
in Figure \ref{fig:grb_200412b}) could arise from one of two physical phenomena.
The first option is that this is the ``jet break'', a geometric effect wherein the jet has cooled
and decelerated enough so that its relativistic beaming angle 1/$\Gamma$ is greater than $\theta_j$, the 
half-opening angle of the conical jet formed by the initial ejecta \citep{rhoads_jet_break}.
This value could be calculated if we were able to obtain constraints for $z$, 
$E_{\rm iso}$, $n_0$, and $\eta$. The other option is that the 
synchrotron cooling spectral break frequency $\nu_c$ is passing 
through the TESS bandpass, causing a steepening of the observed power-law slope; 
in this case, we can either constrain the circum-burst density profile,
or the temporal evolution of other burst parameters, like $E_{\rm iso}$
or the magnetic field fractional energy density $\epsilon_B$.

The jet break is an achromatic effect, meaning that it should be visible in 
observations of the afterglow at all bands. There is no evidence for a break in the
Swift-XRT light curve at the time of the observed break in TESS, as the late-time
Swift-XRT observations (at $\sim$\,4\,d post-burst)  agree with the power law 
from earlier times. Late-time observations
(2\,d post-burst) from the Tautenburg Observatory \citep{klose_tautenburg} claim a
steepening of the power-law decay that could be consistent with a jet break. However, 
these observations were only taken in one band (VB), and the purported time of 
jet break is not consistent with the available Swift-XRT data.

With earlier-time Swift-XRT data, we could also constrain the temporal evolution of
the synchrotron cooling break frequency $\nu_c$---assuming that the observed
break in the light curve arises from $\nu_c$ passing through the TESS band.
\citet{sari_piran_narayan} predict that $\nu_c\,\propto\,t^{-0.5}$; however, significant
deviations from this power-law have been observed in several bursts (see, e.g., 
\citealt{racusin_nakedeye_correlation,filgas_2011}). Decay slopes steeper than the
canonical $-0.5$ could arise from either an accelerated stellar wind (which leads
to a steeper, more negative power-law index in the $n(r)$ wind profile), or temporal
evolution of the fractional magnetic energy density $\epsilon_B$.

\subsection{GRB 210204A}
\label{subsec:210204a_physical}
This GRB has been modeled and thoroughly analyzed in \citet{kumar_afterglow_flares}, and the
best-fit physical parameters are reported in their Tables 1 and 2. 
The TESS light curve of this afterglow reveals tentative evidence for a temporal
power-law break. As for GRB 200412B, we discuss the physical origin of this break (utilizing
the best-fit parameters from the broken power-law fit, in the second row of Table \ref{tab:fit_params}),
and attempt to determine whether this represents the synchrotron cooling break or the
geometrical jet break.  We also discuss new insights from the TESS data about this burst's afterglow.


\subsubsection{The temporal power-law break}
\label{subsubsec:210204a_powerlaw}

For a constant-density circum-burst medium, we utilize the power-law indices from Table \ref{tab:fit_params} and
the relationships for the ``ISM'' case from Table 1 of \citet{zhang_meszaros_prospects_power_law_table}
and Table 2 of \citet{huang_040924}. This calculation yields
electron indices of $p = 1.49\pm0.03$ and $p = 2.21\pm0.20$ when 
$\nu_c < \nu_{\rm TESS}$ and $\nu_c > \nu_{\rm TESS}$, respectively. For the wind case,
we find that $p = 0.82\pm0.03$ and $p = 2.24\pm0.20$. Given that these two values must
be consistent with each other before and after $\nu_c = \nu_{\rm TESS}$, the constant
density ISM case seems the most plausible, given that the early-time and late-time
indices have some overlap at the 3-$\sigma$ level. This may be a plausible interpretation,
but our conclusion is low significance.

In order to evaluate the plausibility of the geometrical jet break interpretation (and constrain
$\theta_j$), we need a robust estimate for the circum-burst density $n$, along with our 
best-fit value for $t_b$ \citep{sari_jets, frail_beaming}.
However, the estimate provided by \citeauthor{kumar_afterglow_flares} 
for $\log_{10}(n_0) = -5.67$ appears relatively low 
compared to the typical values for $n_0$ for GRBs presented in 
Table 2 of \citet{panaitescu_kumar_jets}
and Table 1 of \citet{yost_four_grbs}. Such a low value for $n_0$ is more 
consistent with the comparatively less dense environments around short GRBs 
\citep{fong_sgrb_cbm}; 210204A is not a member of this class of GRBs, given its
extended emission. Modeling of the burst that incorporates the TESS data 
alongside the numerous measurements obtained by \citeauthor{kumar_afterglow_flares}---which
is beyond the scope of this work---may
revise this parameter to be more in line with those observed for long GRBs, and thereby 
robustly constrain $\theta_j$.

\subsubsection{The overall light curve}

\citet{kumar_afterglow_flares} modeled the multi-wavelength afterglow of GRB 210204A
from radio to X-ray, and found that the $r$-band light curve should peak $\sim\,4\times10^3$\,s
post-burst (their Figure 7), when the synchrotron frequency
$\nu_m$ passes through this bandpass. However, the TESS light curve does not show a peak at that predicted time. This discrepancy
is likely explained by the fact that \citeauthor{kumar_afterglow_flares} have only one $r$-band
data point at early times, leading to a poorly-constrained early-time light curve.
The stark difference between their {\tt afterglowpy} model and the TESS data highlights
the degeneracies in afterglow modeling and the uncertainties introduced due to limited data.
This example shows that TESS is able to provide crucial information about the early evolution of GRB afterglows;
while multi-wavelength follow-up was extremely useful in identifying
late-time energy injection in this particular burst, characterizing the early afterglow
is necessary to better constrain key burst parameters.


A potential explanation for the lack of an observed peak could be the presence of a relatively 
bright falling reverse shock occurring at times less than $4\times10^3$ seconds post-burst---which 
would dominate over the afterglow rise before their estimated time of peak. As the afterglow
rises and the reverse shock fades, the sum of these could produce the low-significance flattening
seen in TESS.

\subsection{Other GRBs: 210504A and 230307A}

Given that we have best-fit estimates for the time of afterglow peak for
GRBs 210504A and 230307A, we can qualitatively discuss the Lorentz 
factor $\Gamma_0$ for both. 

For 210504A, we also have a redshift value of $z = 2.077$ from 
X-shooter \citep{210504a_redshift}, which corresponds to a luminosity
distance of 16.6\,Gpc\footnote{For this and all future calculations, we assume a typical $\Lambda$CDM 
cosmology \citep{planck_cosmology} as given in {\tt astropy} \citep{astropy_2013,astropy_2018,
astropy_2022}, with $H_0$ = 67.7\,km\,s$^{-1}$\,Mpc$^{-1}$, $\Omega_m$ = 0.310.}; 
utilizing the reported burst fluence from \citet{swift_bat_210504a} and our best-fit value for
$t_b$ yields an estimate for E$_{\rm iso}$ of $3\times10^{52}$\,erg.
The late $t_{\rm peak}$---especially compared to that of GRB 200412B---suggests a 
lower $\Gamma$ of around 40, implying the launching of a mildly relativistic jet. 

We can perform a similar analysis for GRB 230307A, and compare the results
to those published in \citet{levan_230307a}. This burst is particularly interesting,
as there is evidence that this GRB may arise from a binary neutron star merger
at redshift $z = 0.065$, initially reported by \citet{levan_gcn_jwst_kn} and 
\citet{bulla_gcn_kn}. Our results show
a best-fit afterglow peak time of $\sim7.7\times10^3$\,s. If we adopt the isotropic
energy value and circum-burst densities from \citeauthor{levan_230307a} ($\log_{10}(E_{\rm iso})
 = 51.29$, $\log_{10}(n_0) = -0.62$), and a fiducial $\eta_{0.2} = 1$ as in \citet{molinari_lorentz_fireball},
we find a Lorentz factor of $\sim$25. Given this burst's distance from
its purported host galaxy, we would expect a considerably less dense circum-burst medium,
with a density in line with the values in Table 4 of \citet{fong_sgrb_cbm};
assuming $n_0 = 0.01$\,cm$^{-2}$ increases $\Gamma_0$ to 35.
The difference between the two estimates for $\Gamma_0$ (ours and that from
\citeauthor{levan_230307a})
is most likely attributable to our simplistic heuristic calculation,
compared to their detailed multiwavelength physical modeling. However,
their posterior distribution (Fig. 12 in \citeauthor{levan_230307a})
provides relatively poor constraints on the value of $\Gamma_0$, and they also
sampled $\Gamma_0$ in logarithmic space between $10^2$ and $10^4$.
If this GRB were in fact located
further away, in a higher-redshift galaxy at $z \sim 3.8$ (a possibility discussed 
in \citealt{levan_230307a}), it would have an $E_{\rm iso}$ of 10$^{56}$\,erg
and $\Gamma_0\sim$160.

Another possibility is that the peak observed in the afterglow of 230307A arises
from other late-time central engine activity (such as a flare) rather than the forward shock 
of the afterglow itself. This would suggest that the real afterglow peak occurred between the end of
the gamma-ray emission and the end of the FFI with prompt emission, implying a very rapidly rising 
and falling afterglow---so 
$t_{\rm peak} \lesssim 70$\,s. For the nearby binary neutron star merger
explanation, this would yield $\Gamma_0 \sim 140$; for a faraway collapsar,
we find $\Gamma_0 \gtrsim 920$.

Finally, we highlight the similarity of this light curve to the optical emission 
associated with GRB 090727 (shown in Figure 2 of \citealt{kopac_prompt_sample_2013}), 
where there is a sharp rise and fall at early times, followed by a more gradual rise
and decay for the afterglow.
In that case, \citeauthor{kopac_prompt_sample_2013} rule out 
a reverse external shock origin for the prompt flash and assert that
it is likely caused by an internal shock, where the two 
colliding shells have very different Lorentz factors. However, their
model does not require that the optical and gamma-ray emission originate
from the same region; consequently, the two emission profiles would not
necessarily be temporally correlated. This could suggest that the optical
emission may arise a region such as
the GRB photosphere, rather than from the jet.

TESS did not observe any features in the afterglows for these two bursts
apart from the peak; the cooling and/or jet breaks likely occurred well after 
it faded beyond detectability, preventing us from meaningifully 
constraining $p$ and $\theta_j$.





\section{Discussion}
\label{sec:disc}
We have presented optical light curves for eight GRBs observed by TESS;
these results demonstrate TESS's ability to
obtain well-sampled, early-time optical observations of GRBs. 
Prompt emission was clearly detected at the 3-$\sigma$ level in four GRBs in our sample,
while a flux excess around the time of trigger that
could be consistent with prompt or reverse shock emission was found in two of the
other bursts. The TESS data also allows us to constrain key physical properties 
of the burst---including the Lorentz factor $\Gamma_0$ 
and the electron index $p$. Most importantly, 
observations with TESS come with no opportunity cost, unlike concerted ground-based 
target-of-opportunity follow-up campaigns.
In this section, we estimate GRB detection rates in TESS and 
highlight the utility of TESS for GRB science.

\subsection{Rate estimates and detectability}

Given that the observed distribution of GRBs is 
isotropic on the sky \citep{batse_isotropic}, and TESS observes a large fraction
of the entire sky every two years, we would expect many GRBs 
to fall into the TESS field of view. However, in 
practice, FOVs that include the Milky Way and/or periods of high scattered light
will affect the rate of GRBs with a detectable signal in TESS. 

\citet{smith_191016a} predicted that TESS would be able to detect 
approximately 1 afterglow per year for GRBs that were also detected by
high-energy monitors such as Fermi-GBM or Swift-BAT. 
This value is likely an underestimate, as TESS may observe previously-unidentified
optical afterglows for poorly-localized GRBs.
Here, we aim to estimate the number of concurrent detections
between Swift-BAT and TESS.
From the start of the TESS mission until the end of 2023 ($\sim\,5.4$\,yr),
there were 344 GRBs detected by BAT. This yields a detection rate of 
$\sim$\,64\,yr$^{-1}$, which needs to be corrected by
TESS's sky coverage (5.5\%) and duty cycle (95\%) in order
to estimate the number of optical detections. From this calculation,
we expect approximately 3--4 Swift-BAT GRBs per year to fall in
the concurrent TESS field of view (roughly 1
every 3 sectors).



The previous calculation only accounts for Swift-BAT GRBs; however, there 
could be many more GRBs in the TESS field of view that were discovered by
Fermi-GBM, which has a much larger effective field-of-view, but 
yields localizations that have uncertainties on the order of tens
of square degrees. If we take the $\sim$1300 bursts observed by 
Fermi-GBM between 2018 July to 2023
September (using the {\tt FERMIGBRST} catalog on HEASARC; 
\citealt{fermi_grb_cat_1,fermi_grb_cat_3, fermi_grb_cat_2,fermi_grb_cat_4}), we 
find that Fermi detects approximately 240\,GRB\,yr$^{-1}$. Correcting this
for the TESS FOV and duty cycle, we find that there could be 12--13 GRBs
per year with localizations that overlap the concurrent TESS FOV, or
approximately 1 per observing sector.

As part of our effort to rapidly identify GRBs falling in the TESS field of view,
we have implemented a listener for Swift-BAT GCN Notices that 
will send us an alert whenever a well-localized GRB occurs within the contemporaneous 
TESS field of view. This code is based in part on the overlap tool from \citet{mo_tess_gw}.
The rate estimates that we have calculated here appear to be roughly consistent
with empirical observations---there were 22 GRBs from Swift-BAT, and 70 from Fermi-GBM, in the concurrent
TESS FOV over the initial 5.4 years of the mission, until 2023 December.
We note that these estimates do not include GRBs that lack a high-energy 
trigger, such as the emerging class of ``orphan afterglows'' (see, e.g., \citealt{perley_2019pim}).



While these estimates are realistic, we note that not every GRB in the TESS
field of view will yield a concurrent optical detection---we only had 9
confident detections from our sample of 22 Swift-XRT bursts.
The remaining thirteen GRBs in our sample of 22 (i.e., those not discussed in 
Section \ref{sec:grb_lcs}) were likely not detected by 
TESS due to a combination of factors. For the specific case of GRB 210222B,
which had multiple early-time afterglow detections at $I\sim16$ reported via
the GCN (e.g., \citealt{210222b_lco_afterglow} and \citealt{perley_210222b_afterglow}),
TESS did not detect it because it occurred during an observing gap.
For GRB\,190422A, the
Galactic dust extinction was very high ($A_V\sim\,4.3$),
attenuating the flux by over a factor of 10 
and likely playing a significant role in the non-detection.
For other GRBs, the simplest explanation for the non-detections is that 
the prompt and afterglow emission peaked below the 3-$\sigma$ detection limits,
as enumerated in Table \ref{tab:tess_grb_info}.

\subsection{Constraints on late-time emission}
\label{sec:late_time_searches}

TESS's continuous monitoring of the sky allows us to establish
constraints on late-time emission from GRBs. Supernova (SN)
emission usually peaks $\gtrsim 20$
days after the GRB (see, e.g., \citealt{galama_sn,klose_four_sne}). Kilonovae,
on the other hand, manifest themselves just a few days 
after the trigger (see, e.g., \citealt{jin_kn_lc,gw170817_paper}),
peaking in the TESS band at $\sim\,1$ day post-burst
(Figs. 6 and 8 of \citealt{mo_tess_gw}).
If a GRB is sufficiently nearby and bright in TESS, and falls within the TESS 
field of view during the two sectors ($\sim$60 days) after the burst, we can 
place constraints on fainter late-time emission. 

For all the GRBs in our sample, we visually searched light curves from 2--60\,d after
the burst and found no signals down a limiting magnitude of 18--19. Some of these
non-detections were caused by either TESS's field of view in subsequent sectors shifting away 
from the burst's location, or a significant increase in backgrounds from
scattered light due to the relative orientations of TESS, the Earth,
and the Moon during its orbit.

We can establish a constraint on the redshift of GRB200412B due to 
the lack of detection from any associated SN.
Assuming that a putative GRB-SN is similar to SN1998bw (associated with 
GRB980425)\footnote{This assumption is fairly common in studies of GRB-SNe; see, e.g.,
\citet{srinivasaragavan_boat}.}, it would likely peak at an
$I$-band absolute magnitude of $\sim$ $-$19.2 \citep{galama_sn}. 
This assumption also agrees with the empirical distribution of
GRB-SNe absolute magnitudes from \citet{richardson_sn_dist}.
We binned the data from Sectors 24 and 25 to 6\,h due to the comparatively slow evolution of supernovae,
and to improve the 3-$\sigma$ detection limit to $\sim$\,20.5. Given our non-detection of
any excess emission in Sectors 24 and 25---the two immediately following the burst---
this means that any associated SN must have occurred further than $D_L=870$\,Mpc, so the
GRB must have been located at $z > 0.18$.
Other GRB-SNe have been found to peak at around $-18$ \citep{prentice_stripped_SN};
this value would weaken the distance constraint to $D_L\,\gtrsim\,500$\,Mpc ($z\gtrsim0.11$).

\subsection{Other uses for TESS in GRB science}
\label{subsec:other_uses}

TESS's excellent temporal coverage of GRB afterglows and its
weekly downlink schedule will allow us
to constrain physical parameters that could inform further
multi-wavelength follow-up efforts. As was shown in Section \ref{sec:physical_params}, 
we are able to estimate the initial bulk Lorentz factor $\Gamma_0$ from the afterglow
peak. TESS observations can also identify
afterglows with unusual behavior and inform further
follow-up efforts through extrapolations of the observed power-law decay in order
to predict the feasibility of planned observations.
Conducting such analyses with TESS also do not rely on
resource-intensive target-of-opportunity triggers for multi-wavelength
follow-up.

TESS can investigate the
relationship of the prompt optical flux to the high-energy spectral
energy distribution (SED; see,
e.g. Figure 4 of \citealt{xin_2023_prompt}). An SED constraint 
from TESS at redder wavelengths would help distinguish
between models for the optical flash, clarifying
the relationship between the prompt high-energy and optical emission 
for a given GRB, including the emission region and 
mechanisms (Jayaraman et al., in prep).

Finally, all searches in TESS for optical counterparts to GRBs have relied 
upon a concurrent high-energy detection by a GRB monitoring satellite 
such as Fermi-GBM or Swift-BAT. However, TESS may also be able to detect
so-called ``orphan afterglows'' (which lack a corresponding gamma-ray 
trigger) and ``dirty fireballs'' (transients 
whose relativistic ejecta has too many baryons to produce a GRB). 
Recently, \citet{perley_2019pim} were able to utilize TESS data
to constrain the time of explosion of the fast relativistic transient
AT2019pim and establish 
stringent limits on the detectability of possible gamma-ray emission
from this transient.
The detection of more fast relativistic transients in TESS via a blind 
search (Jayaraman et al., in prep) could trigger timely multi-wavelength 
follow-up, especially with the current cadence of weekly downlinks.

\section{Conclusions and Future Work}

In this work, we present TESS light curves of eight well-localized GRBs
with evidence for prompt optical and/or afterglow emission. 
We constrain physical parameters for four of these bursts using TESS
light curves, and account for the 
TESS cosmic-ray mitigation strategy in order to constrain the brightness
of any associated prompt optical flashes. We 
also briefly discuss the possibilities for TESS to detect late-time emission
from GRBs, and highlight that it is one of the few
observatories capable of detecting both prompt and afterglow emission
from GRBs. 

Modifications to TESS's operations during its Extended Mission 2 have 
enhanced its ability to detect and disseminate information about observations
of prompt optical emission more rapidly, as well as significantly improved
our ability to distinguish between prompt and afterglow 
emission---particularly in the case of GRB 230307A.
Moreover, given that the
weekly downlink schedule now enables the rapid 
creation and public dissemination of the TICA FFIs, we are able to 
identify GRB signatures in TESS within days---considerably earlier than
in previous phases of the mission.

We also note that TESS's large sky coverage, combined with the isotropic
distribution of GRBs on the sky, means that there will be at least a few
GRBs discovered in TESS's field of view each year. This detection rate is
aided by the fact that
TESS has limited downtime---it is neither subject to
the constraints of ground-based observatories (e.g., day-night cycles and weather), 
nor those of space-based observatories in low-earth orbit (e.g., Earth occultation 
and the South Atlantic Anomaly).\footnote{TESS is in a unique, elliptical high-earth 
orbit that places it in a 2:1 resonance with the Moon.} 

Given our predicted rate of up to 10 detections per year
of optical counterparts to GRBs in TESS, we can expect
to identify many more as the mission proceeds. We also
plan to utilize TESS to follow up on GRBs and other relativistic fast transients detected
by the recently-launched Einstein Probe \citep{einstein_probe_2018,einstein_probe_2022},
which is sensitive to GRBs throughout a larger parameter space, including at
higher redshift and in softer bands
(e.g., \citealt{liu_ep,levan_ep}).
Upcoming missions, including the Space Variable Objects Monitor \citep{svom_paper},
will significantly increase
the number of GRBs detected and rapidly localized. TESS clearly fulfills a unique 
role in detecting and characterizing GRBs, and will serve as a useful
complement to other optical large-scale
surveys, such as the ongoing Zwicky 
Transient Facility \citep{bellm_ztf}
and the upcoming Legacy Survey of Space and Time at the Vera Rubin Observatory
\citep{lsst_survey}.
\section*{Acknowledgments}

We thank the anonymous referee for their comments and feedback, which greatly
strengthened the resulting paper.
The authors thank Youdong Hu for information about
BOOTES observations of the afterglow of GRB 220623A. 
RJ also thanks Fr\'ed\'eric Daigne for insightful discussions 
about models for GRB prompt and precursor emission, 
Jochen Greiner for information about GRB catalogs, and Te
Han for discussions regarding photometric techniques to correct
for contaminated sources.

The authors would also like to thank the staff at the Mikulski
Archive for Space Telescopes (MAST), especially Hannah Lewis, Julie Imig, 
Travis Berger, and Scott Fleming, who
assist with the rapid ingest and dissemination of the early-release 
TICA data products.

This paper includes data collected by the TESS mission.  
Funding for TESS  is provided by the NASA Explorer Program.
The TICA data used in this work was 
obtained from MAST
(\dataset[10.17909/t9-9j8c-7d30]{https://dx.doi.org/10.17909/t9-9j8c-7d30}),
hosted by the Space Telescope Science Institute (STScI).
STScI is operated by the Association of Universities for 
Research in Astronomy, Inc., under NASA contract NAS 5–26555.

This work has made use of data from the
UK Swift Science Data Centre at the 
University of Leicester; Swift and
Fermi data provided by the High Energy Astrophysics Science Archive Research 
Center (HEASARC), a service of the Astrophysics Science Division at NASA/GSFC;
and data from the NASA/IPAC Extragalactic Database 
(\dataset[10.26131/IRSA537]{https://dx.doi.org/10.26131/IRSA537}), which is funded 
by the National Aeronautics and Space Administration and operated by the 
California Institute of Technology.


\facilities{TESS, Swift, Fermi}

\software{\texttt{astropy} \citep{astropy_2013,astropy_2018,astropy_2022},
         \texttt{lightkurve} \citep{lightkurve},
         \texttt{matplotlib} \citep{matplotlib},
         \texttt{numpy} \citep{numpy},
         \texttt{scipy} \citep{scipy},
         \texttt{tess-reduce} \citep{tessreduce},
         Fermi-GBM Data Tools \citep{GbmDataTools},
         SpiceyPy/SPICE \citep{spiceypy, NAIF-spice, NAIF-spice-geometry},
         \texttt{swift\_too} (\url{https://www.swift.psu.edu/too_api/}),
         \texttt{synphot} \citep{synphot}
}


\bibliography{tess_grbs}{}
\bibliographystyle{aasjournal}

\appendix
\section{Accounting for Cosmic-Ray Mitigation in TESS}
\label{app:crm}

The TESS detectors read out every 2.0\,s. In order to construct a full-frame image, 
TESS co-adds these 2-second exposures to reach the full cadence of 200\,s, 600\,s, or 1800\,s.
Cosmic rays are rejected during this co-adding process as follows:
The two-second images are examined in sets of 10 (corresponding to a 20s time interval), and
the highest and lowest values for each pixel are discarded, and the remaining 8 values
are summed.  The effective exposure time is reduced by 20\%, but the effect of cosmic
rays is significantly mitigated. Further information about the CRM can be found in 
Section 5.1 of the TESS Instrument
Handbook (\url{https://archive.stsci.edu/files/live/sites/mast/files/home/missions-and-data/active-missions/tess/_documents/TESS_Instrument_Handbook_v0.1.pdf}).

This procedure also affects astrophysical signals that vary on timescales of 10 seconds or less, 
such as the prompt optical emission from GRBs. 
In these cases, TESS's CRM strategy could remove 2\,s cadences that contain signals from the prompt GRB 
emission, leading to an underestimate of the total burst fluence.

We estimated the effect of the TESS CRM technique on the observed optical flux 
from the GRBs in our sample, in order to constrain the 
brightness of any prompt optical emission. This portion of our analysis relies upon
the assumption that the high-energy light curve (from 15--350 keV) is correlated with
the observed prompt optical light curve; this behavior was observed by
\citet{vestrand_2005_prompt,vestrand_130427a}, \citet{racusin_nakedeye_correlation}, 
and a subset of the sample of GRBs presented in \citet{oganesyan_sample_synchrotron}.
It has also been suggested that both the optical and gamma-ray emission may originate 
from a common region \citep{racusin_nakedeye_correlation}. However, other studies (e.g., 
\citealt{sari_piran_990123,060111B_prompt,yost_2007_grb_prompt,gruber_091024}) did not find such a 
correlation between the optical emission and the high-energy prompt emission.
Whether or not such a correlation exists, the fluence detected by TESS for any
prompt counterpart would still
represent a lower limit, due to the cosmic ray mitigation strategy removing flux. The effect
of TESS CRM is even larger if the prompt optical emission consists of several peaks
and spans multiple 20-s clipping bins, as
there will be two or more clipped peaks throughout the period of extended emission.
This could have occurred in the detection of prompt emission for GRB 210204A, where 
the emission lasted for hundreds of seconds.

\begin{figure}
    \centering
    \includegraphics[width=\linewidth]{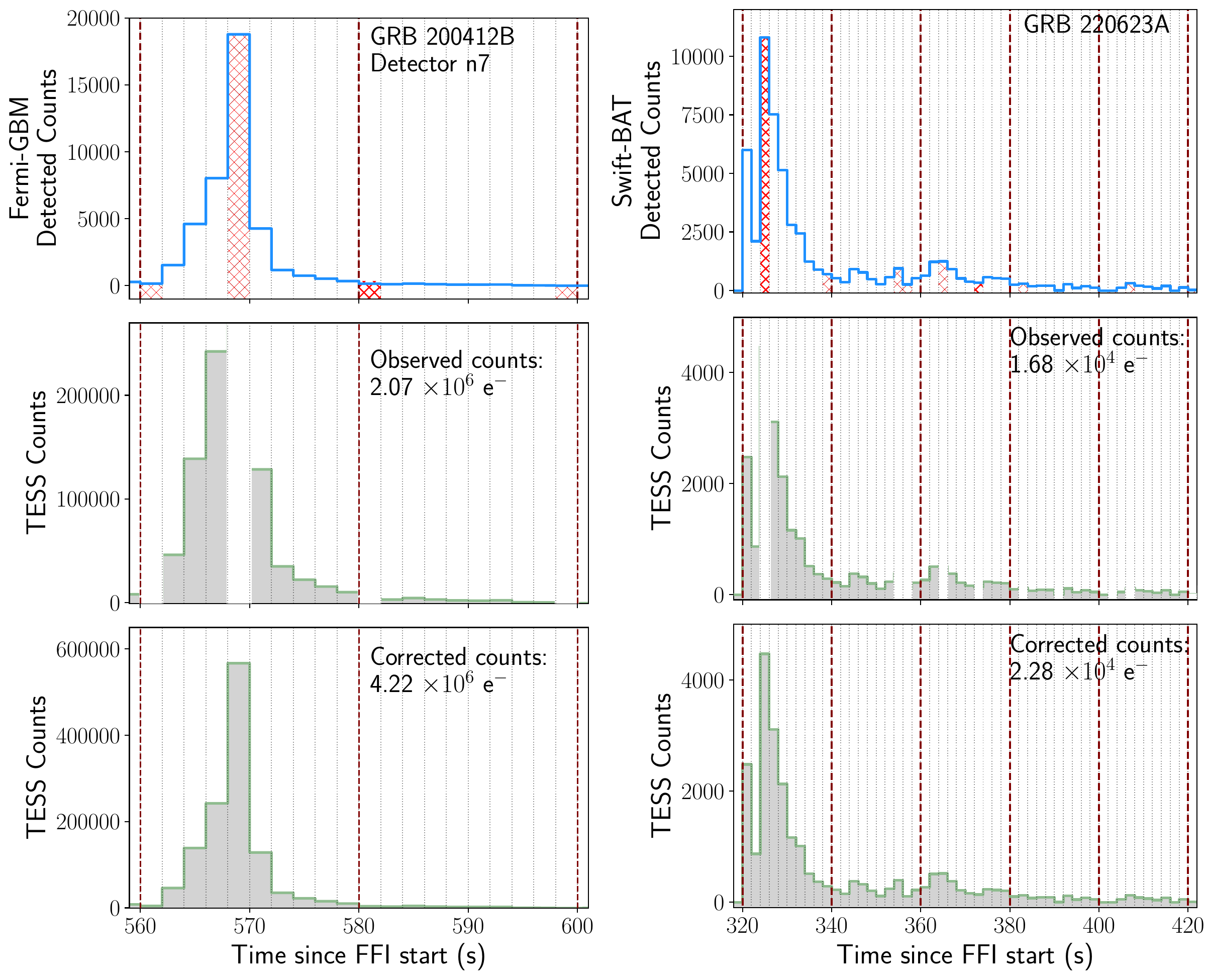}
    \caption{Visual representation of our cosmic-ray mitigation (CRM) correction 
    strategy, as applied to GRBs 200412B and 220623A. The latter has just a prompt
    detection in TESS, while the former has an afterglow and a likely prompt component
    detected (see Section \ref{sec:grb_lcs} for details). We assume that the prompt 
    optical flux tracks the high-energy flux, as in \citet{vestrand_2005_prompt,vestrand_130427a},
    and \citet{racusin_nakedeye_correlation}. 
    The dotted lines show the 2-s TESS sub-exposure boundaries, and the thicker maroon
    dashed lines show the 20-s CRM 
    clipping intervals (see Appendix \ref{app:crm} for details). 
    \textbf{Top}: The Fermi-GBM light curve at 15--350 keV for GRB 200412B (left), and
    the Swift-BAT light curve at 15--350 keV for GRB 220623A (right), binned to the 2-s 
    TESS exposures for the FFI during the prompt emission. The red cross-hatches signify
    the times of maximum and minimum detected counts in each 20-s interval. Assuming
    the optical flux matches the gamma-ray light curve, these intervals would be 
    clipped by the TESS CRM algorithm.
    \textbf{Middle}: The high-energy light curve, scaled to match the 
    observed TESS counts.
    The cadences with the maximum and minimum number of detected counts for each 
    20-s interval are masked.
    The shaded gray area integrates to the total number of counts from the TESS
    light curve (annotation in upper right). For GRB 200412B, this value has been 
    corrected for the extrapolated afterglow contribution.
    \textbf{Bottom}: The TESS light curve for both bursts, with the clipped intervals 
    re-inserted. Over half the optical flux may have been clipped for
    GRB 200412B, while $\sim$25\% was clipped for GRB 220623A.}
    \label{fig:crm_clipping_example}
\end{figure}

Figure \ref{fig:crm_clipping_example} provides a visual representation of how we 
account for the on-board CRM algorithm. 
For this analysis, we assumed the optical prompt emission matched the light curves 
in the 15--350\,keV band; this matches the energy 
regimes from \citet{vestrand_2005_prompt,vestrand_130427a} and \citet{racusin_nakedeye_correlation}. 
First, we calculated the times of the TESS 2-second 
sub-exposures and the 20-second clipping blocks using the {\tt TSTART} and {\tt 
TSTOP} header keywords in the FFIs after correcting to the solar system barycenter,
as described in Section \ref{subsubsec:time_systems}. We then re-binned the high-energy light curve into 
these 2-second bins, and discarded the highest and lowest flux values within each 
20-second block. Next, we integrated the remaining light curve and rescaled it so
that the total counts matched the observed counts from the TESS light curve. Then, 
we added back in the clipped bins from the high-energy light curve, 
rescaled to match the TESS flux scale. Finally, we re-calculated the observed magnitude 
using the corrected counts within various relevant timescales---the 2\,s readout, 
the duration of the GRB (as approximated by $T_{\rm 90}$), and the 
native FFI cadence for that particular sector. These updated values are given in
magnitudes in the discussion for each GRB in Section \ref{sec:grb_lcs}, and 
in counts in Table \ref{tab:flux_ratios}.

This strategy for handling CRM relies on the assumption that the clipped bins are 
the same for each pixel that has a signal from the GRB. The clipped
bins may not be the same for all the pixels within the photometric aperture. 
However, the source counts per pixel near the core of the PSF is 10 times larger 
than the total noise per pixel for the faintest prompt detection in our sample, 
GRB 230903A. Because the wings of the PSF contribute little to the integrated flux, 
assuming that the clipped bins are the same for each pixel for the 
optical flash is reasonable. 

\end{document}